\documentstyle[12pt,epsf,epsfig,a4]{article}
\begin{document}
\begin {titlepage}
\begin{flushleft}
FSUJ TPI QO-19/96
\end{flushleft}
\begin{flushright}
November, 1996
\end{flushright}
\vspace{20mm}
\begin{center}
{\Large \bf Schr\"odinger cat-like states by 
conditional measurements on a beam-splitter  } 
\vspace{15mm}
 
{\large \bf M. Dakna, T. Anhut, T.  Opatrn\'{y}, 
L. Kn\"oll, and D.--G. Welsch}\\[1.ex]
{\large Friedrich-Schiller-Universit\"at Jena}\\[0.5 ex]
{\large Theoretisch-Physikalisches Institut}\\[0.5 ex]
{\large Max-Wien Platz 1, D-07743 Jena, Germany}
\vspace{25mm}
\end{center}
\begin{center}
\bf{Abstract}
\end{center}
A scheme for generating Schr\"{o}dinger cat-like states
of a single-mode optical field by means of conditional measurement
is proposed. Feeding into a beam splitter a squeezed vacuum
and counting the photons in one of the output channels, the
conditional states in the other output channel exhibit a number
of properties that are very similar to those of superpositions of two
coherent states with opposite phases.
We present analytical and numerical results for the
photon-number and quadrature-component distributions of the
conditional states and their Wigner and Husimi functions. Further, we
discuss the effect of realistic photocounting on the states.
\end{titlepage}
\renewcommand {\thepage} {\arabic{page}}
\setcounter {page} {2}
\section{Introduction}
 \label{sec1}

It is well known that according to the basic-theoretical principles of
quantum-mechanics the superposition of macroscopically distinguishable
quantum states can give rise to quantum interferences such that the
resulting states are highly nonclassical. Schr\"{o}dinger
\cite{Schroedinger1} illustrated this phenomenon by a gedankenexperiment
to get a cat into a superposition of a live and a dead cat.
A number of systems have been studied with the aim of the
realization of Schr\"{o}dinger cat-like states, where the ``cat'' is
typically a mesoscopic system that has both microscopic and macroscopic
(i.e., classically distinguishable) properties.

In harmonic oscillators typical examples of Schr\"{o}dinger
cat-like states are superpositions of two coherent (i.e., most classical)
states with opposite phases \cite{Yurke1}. The superposition states
exhibit some properties similar to those of simple statistical mixtures,
but they also reveal typical interference features.
Measuring the quadrature-component distribution,
one observes two peaks that change their mutual distance in
dependence on the phase of the quadrature component until
they eventually overlap. In this particular case the difference
between a coherent superposition and a statistical mixture is the
most distinct. Whereas in the former case quantum interferences are
observed which give rise to an oscillatory behavior of the
quadrature-component distribution, in the latter case a single peak
without interference structure is observed. The creation of such
states in realistic experiments is not trivial. Proposals for
preparing vibrations in molecules or crystals in Schr\"{o}dinger
cat-like states have been made \cite{Janszky1,Walmsley1}. Recently,
it has been proposed that Schr\"{o}dinger cat-like states of a single
harmonically bound (trapped) atom can be produced by appropriately
driving the atom \cite{Vogel1}, and experiments have successfully
been performed \cite{Monroe1}. The possibility of preparing a
harmonically bound atom in a superposition of two coherent squeezed
states with opposite phases has also been studied \cite{Nieto1}.

Several proposals have been made to prepare a single-mode
radiation field (which also corresponds to a harmonic oscillator)
in a Schr\"{o}dinger cat-like state (see, e.g., \cite{Buzek1} and
references therein). In particular, it has been proposed that
conditional measurements may  advantageously be used to realize
such states. It is well known that if a quantity of a subsystem of a
correlated two-part system prepared in some entangled state is
measured, the state of the other subsystem ``collapses'' to a
particular state. To produce conditional states of the type of
Schr\"{o}dinger cat-like states, the use of a scheme for
optical back-action-evading measurement in nonlinear media
\cite{LaPorta1} was suggested \cite{Song1}. The calculations
show that when the photon number of the readout mode is measured,
then a superposition of macroscopically distinguishable quantum states
is generated in the signal mode \cite{Song1}. To improve the
scheme, it was proposed that a squeezed vacuum at the
signal frequency is injected instead of amplifying the signal after
back-action-evading measurement \cite{Yurke2}. Recently,
a modification of this scheme was studied, with special
emphasis on the experimental feasibilities with current
technologies \cite{Tombesi1,Tombesi2}.

In this paper we show that Schr\"{o}dinger cat-like states can
already be obtained using a simple beam-splitter scheme for
a conditional measurement of the type considered recently
in \cite{Ban1,Ban2}. The calculations show that
when a squeezed vacuum is injected in one
of the input channels (the second input channel being unused)
and the photon number of the mode in one of the output
channels is measured, then the mode in the other output channel
is prepared in a conditional state that has the typical
features of a Schr\"{o}dinger cat-like state. In particular,
the conditional states can be regarded as superpositions of two
quantum states that are well localized in the phase space and bear
a strong resemblance to squeezed coherent states. To demonstrate
this, we analyse the states in terms of the photon-number
and quadrature-component distributions and the Wigner and Husimi
functions. We further discuss possible modifications of the states in
realistic multichannel photon detection.

The paper is organized as follows. In Sec.~\ref{sec2} the conditional
states are derived, and in Sec.~\ref{sec3} their properties are
discussed. Section \ref{sec4} is devoted to a realistic detection
scheme. Finally, a summary and some concluding remarks are given in
Sec.~\ref{sec5}.

%%%%%%%%%%%%%%%%%%%%%%%%%%%%%%%%%%%%%%%%%%%%%%%%%%%%%%%%%%%%%%%%%%%%%%%%

\section{Basis equations}
\label{sec2}

Let us consider a lossless beam splitter and assume that the
input fields can be regarded as being effectively single mode fields,
with photon destruction and creation operators $\hat{a}_{k}$ and
$\hat{a}^{\dagger}_{k}$, respectively ($k\!=\!1,2$).
The photon destruction and creation operators
of the output modes $\hat{b}_{k}$ and $\hat{b}^{\dagger}_{k}$,
respectively, can then be obtained using the well-known input--output
relations
\begin{equation}
\hat{b}_{k} = \sum_{k'=1}^{2} T_{k,k'} \, \hat{a}_{k'},
\label{1.01}
\end{equation}
where
\begin{eqnarray}
(T_{k,k'}) = e^{i\varphi_0}
\left(\begin{array}{cc}\cos\theta
\;e^{i\varphi_{R}}&\sin\theta\;e^{i\varphi_{R}}\\
-\sin\theta\;e^{-i\varphi_{R}}&\cos\theta\;e^{-i\varphi_{T}}
\end{array}\right)
\label{1.02}
\end{eqnarray}
is a SU(2) matrix whose elements are given by the complex
transmittances and reflectances of the beam splitter from the
two sides. Equation (\ref{1.01}) corresponds to a unitary transformation
of the operators in the Heisenberg picture, $\hat{b}_{k}\!=\!\hat{V}
\hat{a}_{k} \hat{V}^{\dagger}$ ($k\!=\!1,2$).
Equivalently, the Schr\"{o}dinger picture
can be used, in which the photonic operators are left unchanged and the
density operator is transformed such that the output-state density
operator $\hat{\varrho}_{\rm out}$ is obtained from the input-state
density operator $\hat{\varrho}_{\rm in}$ as
\begin{equation}
\hat{\varrho}_{\rm out}
= \hat{V}^{\dagger} \hat \varrho_{\rm in}  \hat{V}.
\label{1.03}
\end{equation}
The operator $\hat{V}$ can be given by \cite{Yurke3,Campos1}
\begin{equation}
\hat{V} =
e^{-i(\varphi_{T}-\varphi_{R}) \hat{L}_{3}}
\, e^{-2i\theta \hat{L}_{2}}
\, e^{-i(\varphi_{T}+\varphi_{R}) \hat{L}_{3}},
\label{1.03a}
\end{equation}
where
\begin{equation}
\hat{L}_{2} = \textstyle\frac{1}{2}(\hat{a}_{1}^\dagger\hat{a}_{2}
-\hat{a}_{2}^\dagger\hat{a}_{1}), \quad
\hat{L}_{3} = \textstyle\frac{1}{2}(\hat{a}_{1}^\dagger\hat{a}_{1}
-\hat{a}_{2}^\dagger\hat{a}_{2}),
\label{1.03b}
\end{equation}
and $\varphi_{0}\!=\!0$ (note that $\varphi_{0}$ is a global phase
factor that may be omitted without loss of generality).

Now, let us assume that the mode in the first input channel
is prepared in a state described by a density operator
$\hat{\varrho}_{{\rm in}1}$ and the second channel is unused,
so that the input-state density operator reads as
\begin{equation}
\hat \varrho_{\rm in} = \hat \varrho_{{\rm in}1}
\otimes |{\rm vac}_{2}\rangle \langle{\rm vac}_{2}|.
\label{1.04}
\end{equation}
Using Eqs.~(\ref{1.03a}) and (\ref{1.04}), the output-state density
operator (\ref{1.02}) can be given by \cite{Ban1,Ban2}
\begin{eqnarray}
\lefteqn{
\hat \varrho _{\rm out} =
\sum_{n_{2}=0}^{\infty}\sum_{m_{2}=0}^{\infty}
\bigg\{
\frac{e^{-i(m_2-n_2)\varphi_{R}}}
{\sqrt{m_{2}!n_{2}!}}
\left|\frac{R}{T}\right|^{m_{2}+n_{2}}
\!\!e^{i\varphi_{T}\hat{a}^\dagger_{1}\hat{a}_{1}}
}
\nonumber \\  && \hspace{1ex} \times \,
\hat{a}_{1}^{m_{2}}
 |T|^{\hat{a}^\dagger_{1}\hat{a}_{1}}
\hat{\varrho}_{{\rm in}1}
|T|^{\hat{a}_{1}^\dagger\hat{a}_{1}}({\hat{a}_{1}^\dagger})^{n_{2}}
e^{-i\varphi_{T}\hat{a}^\dagger_{1}\hat{a}_{1}}
\otimes |m_{2}\rangle\langle n_{2}|
\bigg\},
\nonumber \\ &&
\label{1.0}
\end{eqnarray}
where $|T|\!=\!\cos\theta$ and $|R|\!=\!\sin\theta$, and
$|n_{k}\rangle$ are the eigen\-states of the photon-number operators
$\hat{a}^{\dagger}_{k}\hat{a}_{k}$. From Eq.~(\ref{1.0}) we see that the
output modes are, in general, highly correlated. When the photon number
of the mode in the second output channel is measured and $m_2$ photons are
detected, then the mode in the first output channel is prepared in a
quantum state whose density operator $\hat{\varrho}_{{\rm out}1}$ reads as
\begin{eqnarray}
\hat{\varrho}_{{\rm out}1}(m_2)
= \frac{\langle m_{2} |\hat{\varrho}_{\rm out} | m_{2} \rangle}
{{\rm Tr}_{1}( \langle m_{2} | \hat{\varrho}_{\rm out} | m_{2} \rangle ) } .
\label{1.1}
\end{eqnarray}
The probability of such an event is given by
\begin{eqnarray}
\lefteqn{
P(m_2)=
{\rm Tr}_{1}( \langle m_{2} | \hat{\varrho}_{\rm out} | m_{2} \rangle )
}
\nonumber \\ && \hspace{1ex}
= \!\! \sum_{n_{1}=m_{2}}^{\infty}
\!\!{n_{1}\choose m_{2}}  \!
\left(1-|T|^2\right)^{m_{2}} |T|^{2(n_{1}-m_{2})}
\langle n_{1} | \hat{\varrho}_{{\rm in}1} | n_{1} \rangle.
\label{1.6}
\end{eqnarray}

In particular, if the first input mode is prepared in a
squeezed vacuum state, we may write
\begin{equation}
\hat{\varrho}_{{\rm in}1}
= \hat{S}(\xi)| {\rm vac}_{1} \rangle
\langle {\rm vac}_{1}  | \hat{S}^{\dagger}(\xi),
\label{1.61}
\end{equation}
where
\begin{eqnarray}
\lefteqn{
\hat{S}(\xi)| {\rm vac}_{1} \rangle =
\exp\!\left\{-\textstyle\frac{1}{2}
\left[(\xi\hat{a}_{1}^{\dagger})^{2} - \xi^{\ast}\hat{a}_{1}^{2}\right]
\right\} | {\rm vac}_{1} \rangle
}
\nonumber \\ & & \hspace{4ex}
=\,(1-|\kappa|^{2})^{\frac{1}{4}}\sum_{n_{1}=0}^{\infty}
\frac{[(2n_{1})!]^{1/2}}{2^{n_{1}}\,n_{1}!}\,
\kappa^{n_{1}}|2n_{1}\rangle,
\label{1.2}
\end{eqnarray}
$\xi\!=\!|\xi|e^{i\varphi_{\xi}}$, $\kappa\!=\!e^{i\varphi_{\xi}}\tanh|\xi|$.
Combining Eqs.~(\ref{1.0}) and (\ref{1.1}) and using Eqs.~(\ref{1.04}),
(\ref{1.61}), and (\ref{1.2}), we derive that
\begin{equation}
\hat{\varrho}_{{\rm out}1}(m_{2})
= | \Psi_{m_{2}} \rangle \big\langle \Psi_{m_{2}} |,
\label{1.3}
\end{equation}
where
\begin{equation}
|\Psi_{m_{2}}\rangle = |\Psi_{m_{2}}(\alpha)\rangle =
\frac{1}{\sqrt{{\cal N}_{m_{2}}}}
\sum_{n_{1}=0}^{\infty} c_{m_{2},n_{1}}(\alpha) \, |n_{1}\rangle,
\label{1.4}
\end{equation}
\begin{eqnarray}
\lefteqn{
c_{m_{2},n_{1}}(\alpha) =
\frac{(n_{1}+m_{2})!}
{\Gamma\!\left[\frac{1}{2}(n_{1}+m_{2})+1\right]\sqrt{n_{1}!}}
}
\nonumber \\ && \hspace{4ex} \times \,
\textstyle\frac{1}{2}\left[1+(-1)^{n_{1}+m_{2}}\right]
\left(\textstyle\frac{1}{2}\alpha\right)^{\frac{1}{2}(n_{1}+m_{2})},
\label{1.5}
\end{eqnarray}
$\alpha\!=\!|\alpha|e^{i\varphi_{\alpha}}$, with
$|\alpha|\!=\!|T|^{2}|\kappa|$ and $\varphi_{\alpha}\!=\!
2\varphi_{T}+\varphi_{\xi}$. In what follows we will restrict
attention to real values of $\alpha$ ($-1\!\leq\!\alpha\!\leq\!1$), i.e.,
$\varphi_{\alpha}\!=\!0,\pi$, since from Eqs.~(\ref{1.4}) and (\ref{1.5})
the effect of other phases $\varphi_{\alpha}$ is simply a rotation in phase
space. Applying the relations (\ref{A.1a}),
(\ref{A.2}), and (\ref{A.3}) in the appendix, the normalization constant
\begin{equation}
{\cal N}_{m_{2}} = \sum_{n_{1}=0}^{\infty} \left|c_{m_{2},n_{1}}\right|^{2}
\label{1.51}
\end{equation}
can be given by
\begin{eqnarray}
\lefteqn{
{\cal N}_{m_{2}}
= \frac{1}{\sqrt{1-\alpha^2}}
\left[\frac{\alpha ^2}{(1-\alpha^2)}\right]^{m_{2}}
}
\nonumber \\ && \hspace{10ex} \times \,
\sum_{k=0}^{\left[\frac{1}{2}m_{2}\right] }
\frac{\left(m_{2}!\right)^2}
{(m_{2}-2k)! \left( k! \right)^{2}(2\alpha)^{2k}},
\label{1.3b}
\end{eqnarray}
where the symbol $[x]$ in the summation upper limit denotes the 
integral part of $x$.
Similarly, using Eqs.~(\ref{1.04}), (\ref{1.61}), and (\ref{1.2})
and applying the relations (\ref{A.1a}), (\ref{A.2}), and
(\ref{A.3}), Eq.~(\ref{1.6}) yields
\begin{eqnarray}
\label{P2}
\lefteqn{
P(m_{2})
= \sqrt{\frac{1-\kappa^2}{1-\alpha^2}}
\left[\frac{\alpha ^{2} (1-|T|^2)}
{|T|^2(1-\alpha^2)}\right]^{m_{2}}
}
\nonumber \\ && \hspace{10ex} \times \,
\sum_{k=0}^{\left[ \frac{1}{2}m_{2} \right]}
\frac{m_{2}!}{(m_{2}-2k)!(k!)^{2}(2\alpha)^{2k}}.
\end{eqnarray}

   From Eqs.~(\ref{1.4}) and (\ref{1.5}) we easily see that when the
detected number of photons in one output channel,
$m_{2}$, is even (odd), then the mode in the other output channel
is prepared in a quantum state $|\Psi_{m_{2}}\rangle$ that contains
only contributions from photon-number states with even (odd) numbers
of photons. This property, which gives rise to oscillations in the
photon-number distribution of the output state $|\Psi_{m_{2}}\rangle$,
obviously reflects the fact that the squeezed vacuum that is fed in
consists of pairs of photons. In particular, when the number of detected
photons, $m_{2}$, is zero, then the output state is again a squeezed vacuum,
but with the parameter $\alpha$ in place of~$\kappa$.

%%%%%%%%%%%%%%%%%%%%%%%%%%%%%%%%%%%%%%%%%%%%%%%%%%%%%%%%%%%%%%%%%%%%%%%

\section{Properties of the conditional states}
\label{sec3}

To study the properties of the conditional states $|\Psi_{m}\rangle
\!=\!|\Psi_{m}(\alpha)\rangle$ in more detail, we will calculate
the photon-number and quadrature-component
distributions and the Wigner and Husimi functions. Further, we
will show that the states $|\Psi_{m}\rangle$ can be represented as
superpositions of two macroscopically distinguishable ``quasicoherent''
squeezed states. For notational convenience we will omit the subscripts $1$
and $2$ introduced above to distinguish between the two output channels.

%%%%%%%%%%%%%%%%%%%%%%%%%%%%%%%%%%%%%%%%%%%%%%%%%%%%%%%%%%%%%%%%%%%%%%%%%

\subsection{Photon-number distribution}
\label{sec34}

\begin{figure}
\centering\epsfig{figure=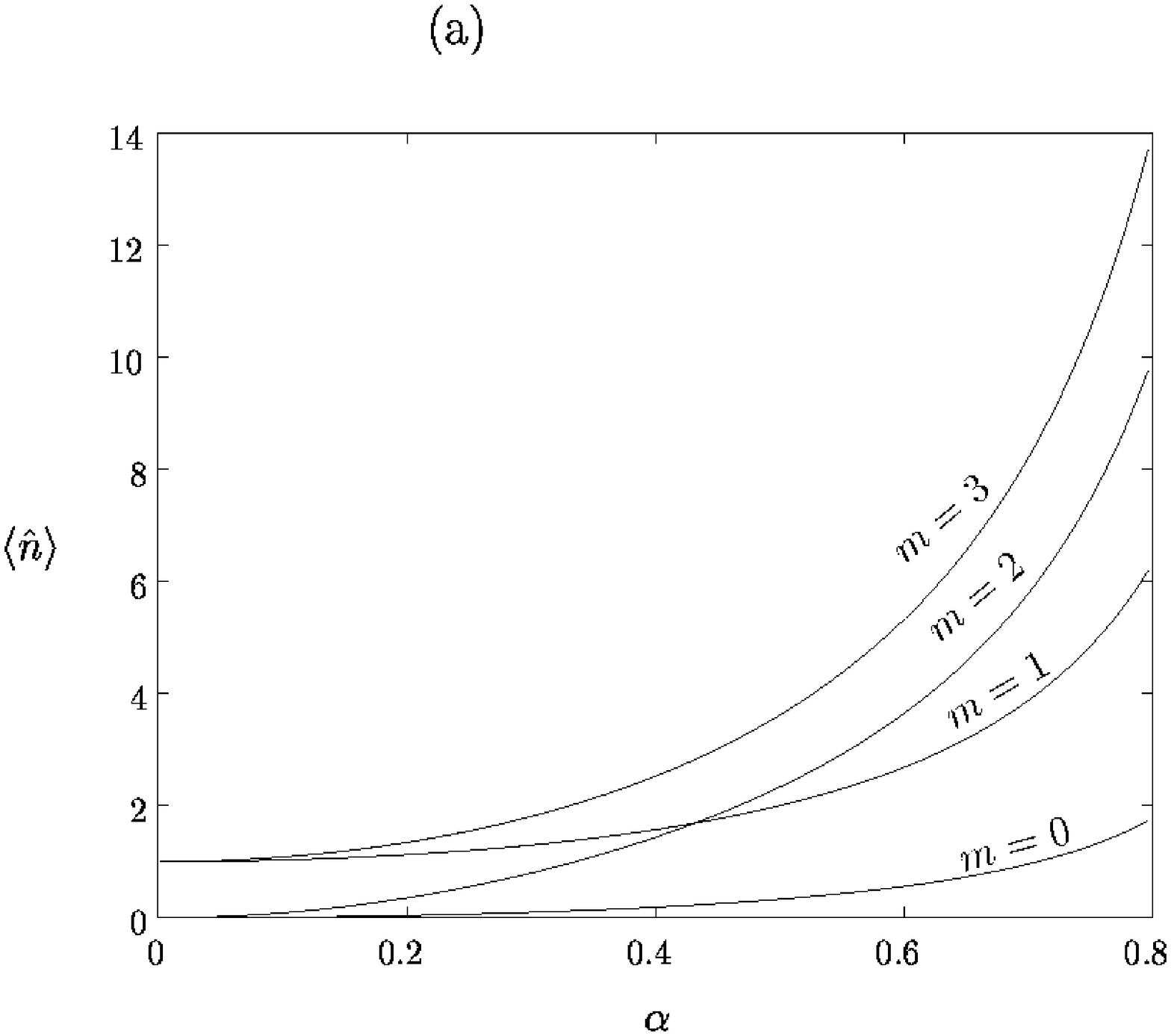,width=0.8\linewidth}
\centering\epsfig{figure=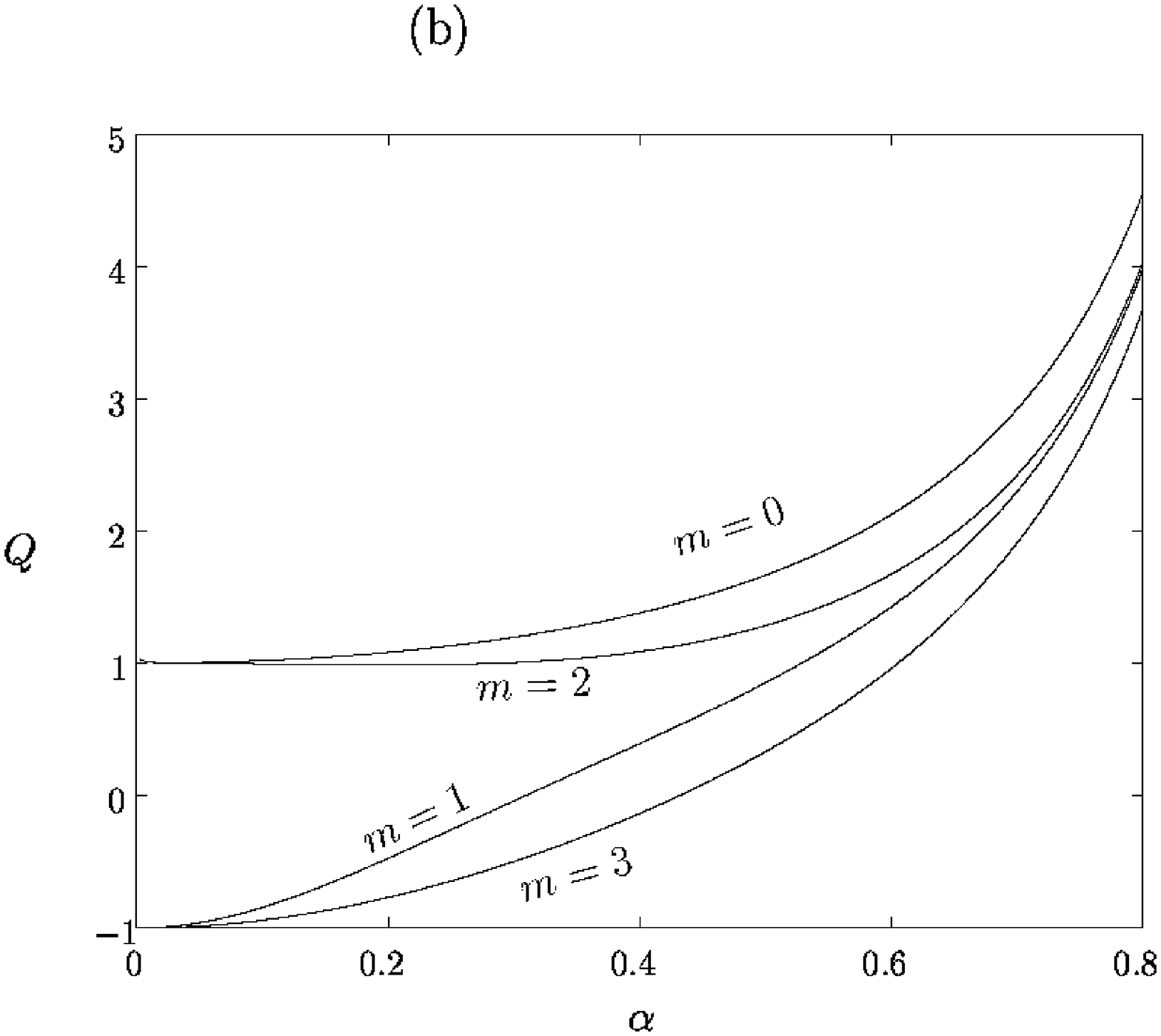,width=0.8\linewidth}
\caption{ \label{fig0}(a) Mean photon number $\langle\hat{n}\rangle$ of the
conditional state $|\Psi_{m}\rangle$ as a function of $\alpha$
for various numbers $m$ of measured photons.
(b) Dependence on $\alpha$ of the Mandel $Q$ parameter for the
same values of $m$ as in (a).  }
\end{figure}

Recalling Eq.~(\ref{1.4}), the photon-number distribution
\begin{equation}
P(n|m) = |\langle\,n|\Psi_m\rangle|^2
\label{n1}
\end{equation}
of a state $|\Psi_{m}\rangle$ reads as
\begin{equation}
P(n|m) = {\cal N}_{m}^{-1} \left| c_{m,n}(\alpha) \right|^{2},
\label{n2}
\end{equation}
where $c_{m,n}(\alpha)$ and ${\cal N}_{m}$ are given in Eqs.~(\ref{1.5})
and (\ref{1.51}), respectively. In particular, the mean photon number
\begin{equation}
\langle\hat{n}\rangle
={\cal N}_{m}^{-1}\sum_{n=0}^{\infty}n|c_n(\alpha,m)|^2
\label{n3}
\end{equation}
can be given by
\begin{eqnarray}
\lefteqn{
\langle\hat{n}\rangle=
\alpha\frac{\partial}{\partial\alpha}
\log\!\left[\frac{{\cal N}_m}{\alpha^m}\right]
}
\nonumber \\ && \hspace{1ex} = \,
\frac{\alpha^2}{1-\alpha^2}+m\frac{1+\alpha^2}{1-\alpha^2}
-2 \sum_{k=0}^{\left[\frac{{1}}{2}m\right] }k\,a_{k,m}
\left[\sum_{k=0}^{\left[\frac{{1}}{2}m\right] }a_{k,m}\right]^{-1}\!\!,
\nonumber \\ &&
\label{n4}
\end{eqnarray}
where $a_{k,m}=(2\alpha)^{-2k}/[(m-2k)!\left( k! \right)^{2}]$.
Examples are shown in Fig.~\ref{fig0}(a). We see that the number of
photons that can be found in $|\Psi_{m}\rangle$ increases with $m$.
This is simply a consequence of the beam splitter transformation.
Since one of the input channels is unused, the mean numbers of
photons in the two output channels are proportional to each other,
the ratio being given by $|T/R|^{2}$. Note that when no photons are
detected, $m\!=\!0$, then $\langle\hat{n}\rangle$ reduces to the mean
number of photons of a squeezed vacuum, $\langle\hat{n}\rangle\!=
\! \alpha^2/(1-\alpha^2)$.

A measure of the deviation of the photon-number distribution from a
Poissonian is the Mandel $Q$ parameter \cite{Mandel1}
\begin{equation}
Q = \frac{\langle\hat{n}^2\rangle-\langle\hat{n}\rangle^2}
{\langle\hat{n}\rangle}-1,
\label{n5}
\end{equation}
which can be given by
\begin{equation}
Q = \frac{\alpha^{2}}{\langle\hat{n}\rangle} \,
\frac{\partial^2}{\partial\alpha^2}
\log\!\left[\frac{{\cal N}_m}{\alpha^m}\right]\\
=\frac{\alpha}{\langle\hat{n}\rangle}\frac{\partial}{\partial\alpha}
\langle\hat{n}\rangle-1 .
\label{n6}
\end{equation}
The dependence on $\alpha$ and $m$ of $Q$ is shown in Fig.~\ref{fig0}(b).
In particular, for even $m$ we find that $Q\!>\!0$ for all values
of $\alpha$, which means that the photon-number statistics of
$|\Psi_{m}\rangle$ is super-Poissonian. For odd $m$
and small values of $|\alpha|$ the statistics become sub-Poissonian 
($Q\!<\!0$). Note that the behavior is typical for
Schr\"odinger cat-like states \cite{Buzek1}.

%%%%%%%%%%%%%%%%%%%%%%%%%%%%%%%%%%%%%%%%%%%%%%%%%%%%%%%%%%%%%%%%%%%%%%%%

\subsection{Quadrature distributions}
\label{sec33}

In order to calculate the conditional quadrature-component distribution
(i.e., the phase-parametrized field-strength distribution)
\begin{equation}
p(x,\varphi|m)=|\langle x,\varphi|\Psi_{m}\rangle|^2,
\label{F1}
\end{equation}
which can be measured in balanced homodyne detection, we first expand
the eigenvectors $|x,\varphi\rangle$ of the quadrature component
\begin{equation}
\hat{x}(\varphi)
= 2^{-\frac{1}{2}}
\left(e^{-i\varphi}\hat{a}+e^{i\varphi}\hat{a}^\dagger\right)
\label{F2}
\end{equation}
in the photon-number basis as \cite{Vogel2}
\begin{equation}
|x,\varphi\rangle=(\pi)^{-\frac{1}{4}}
\exp\!\left(-\textstyle\frac{1}{2}x^2\right)
\sum_{n=0}^{\infty}\frac{e^{in\varphi}}
{\sqrt{2^nn!}} \, {\rm H}_n(x) |n \rangle
\label{F3}
\end{equation}
(H$_{n}$ is the Hermite polynomial). Using  Eqs.~(\ref{1.4}) and
(\ref{F3}) and applying the relations (\ref{A.2}) and (\ref{A.3}),
the conditional quadrature-component distribution (\ref{F1}) reads as
\begin{eqnarray}
\lefteqn{
p(x,\varphi|m)=
\frac{|\alpha |^{m}}{{\cal N}_{m} \sqrt{\pi \Delta^{m+1}} \, 2^{m}}
\exp\!\left( - \frac{1-\alpha ^{2}}{\Delta} \, x^{2} \right)
}
\nonumber \\ && \hspace{15ex} \times \,
\Big |
{\rm H}_m\!\left[
\sqrt{(\alpha e^{i2\varphi}-\alpha^2)/\Delta}\,x
\right]\Big |^2 ,
\label{F4}
\end{eqnarray}
where the abbreviation
\begin{equation}
\Delta = 1 + \alpha^2 - 2 \alpha \cos(2\varphi)
\label{F5}
\end{equation}
has been used.

\begin{figure}
\noindent
\begin{minipage}[b]{0.5\linewidth}
\centering\epsfig{figure=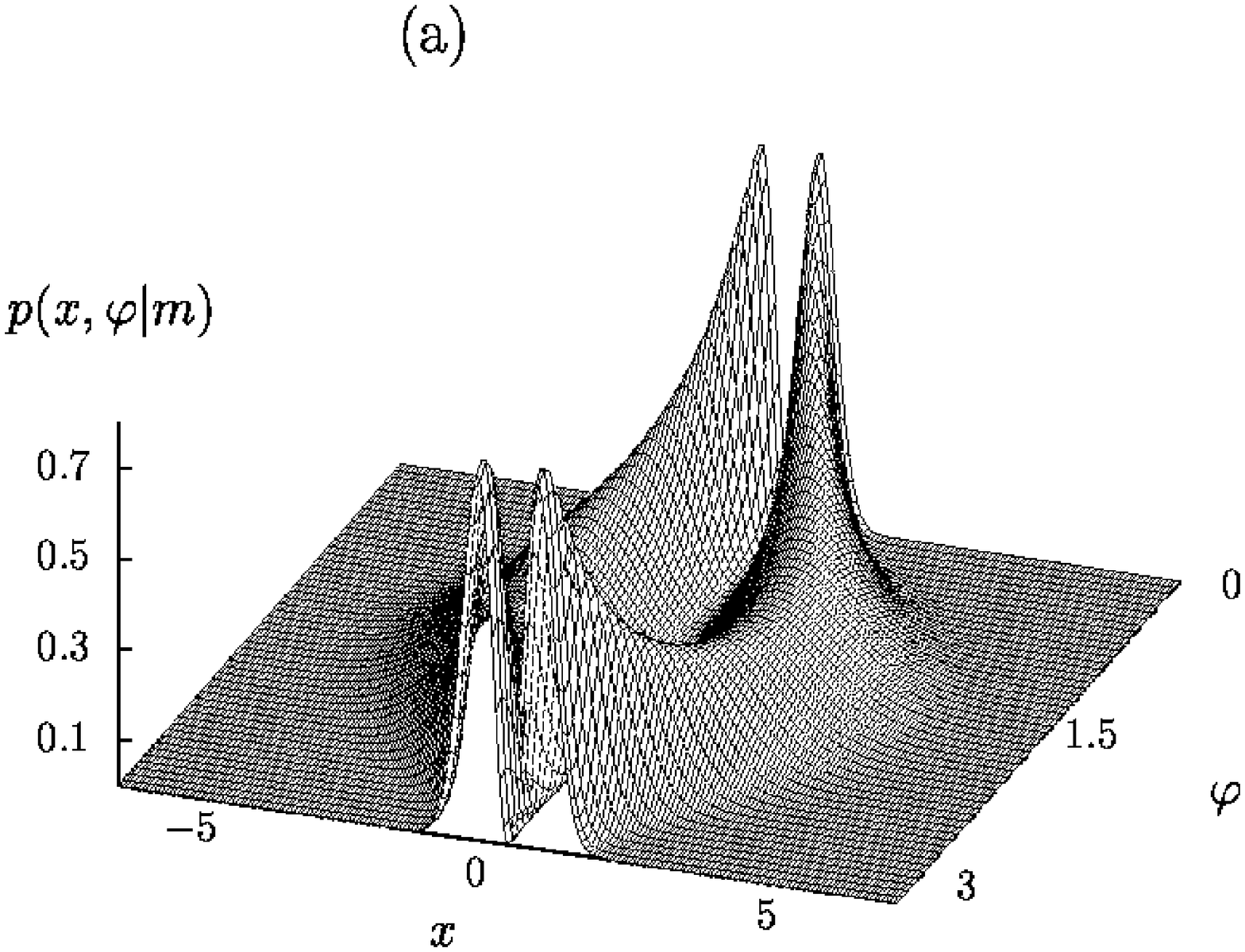,width=\linewidth}
\end{minipage}\hfill
\begin{minipage}[b]{0.5\linewidth}
\centering\epsfig{figure=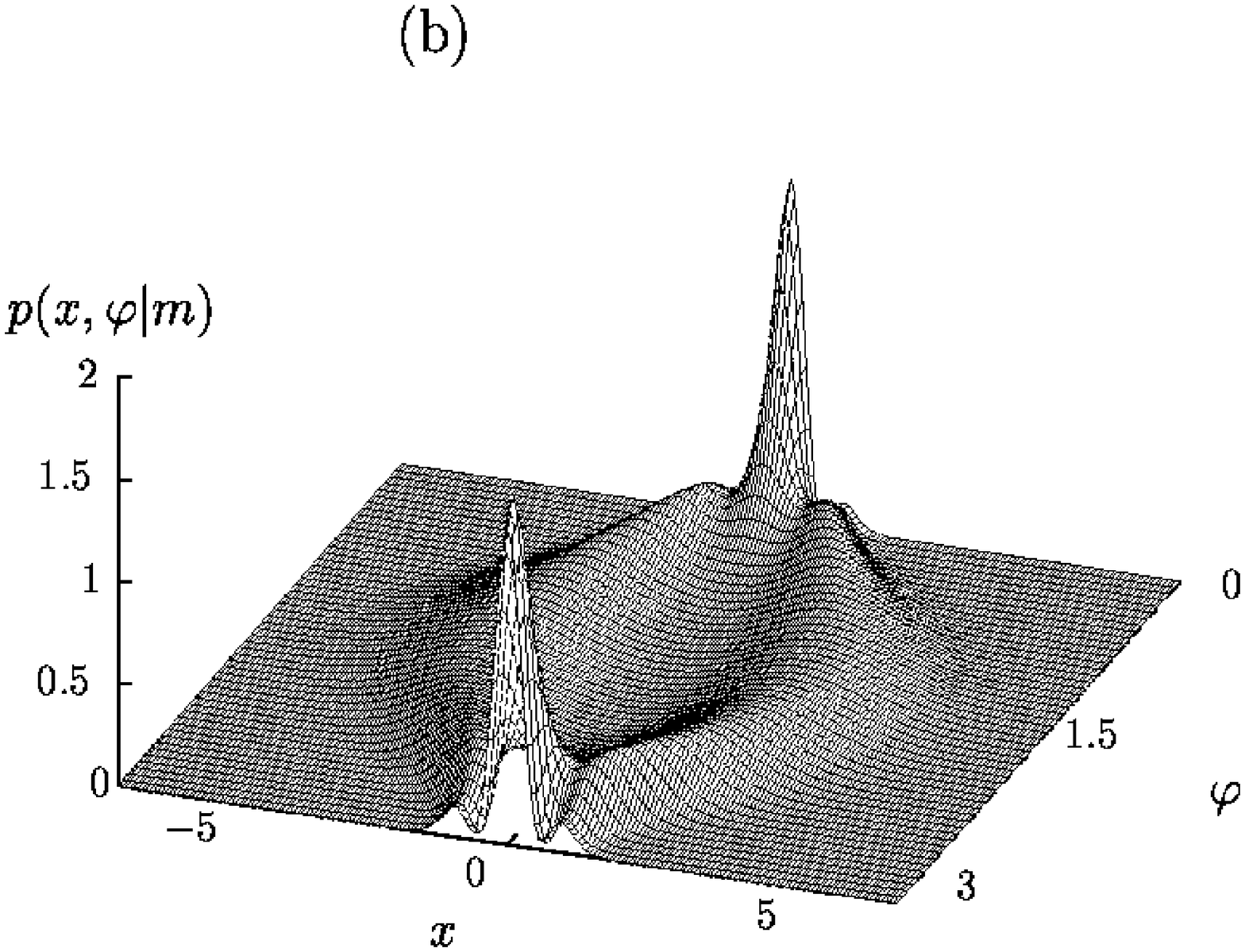,width=\linewidth}
\end{minipage}

~

\vspace{2cm}

~

\begin{minipage}[b]{0.5\linewidth}
\centering\epsfig{figure=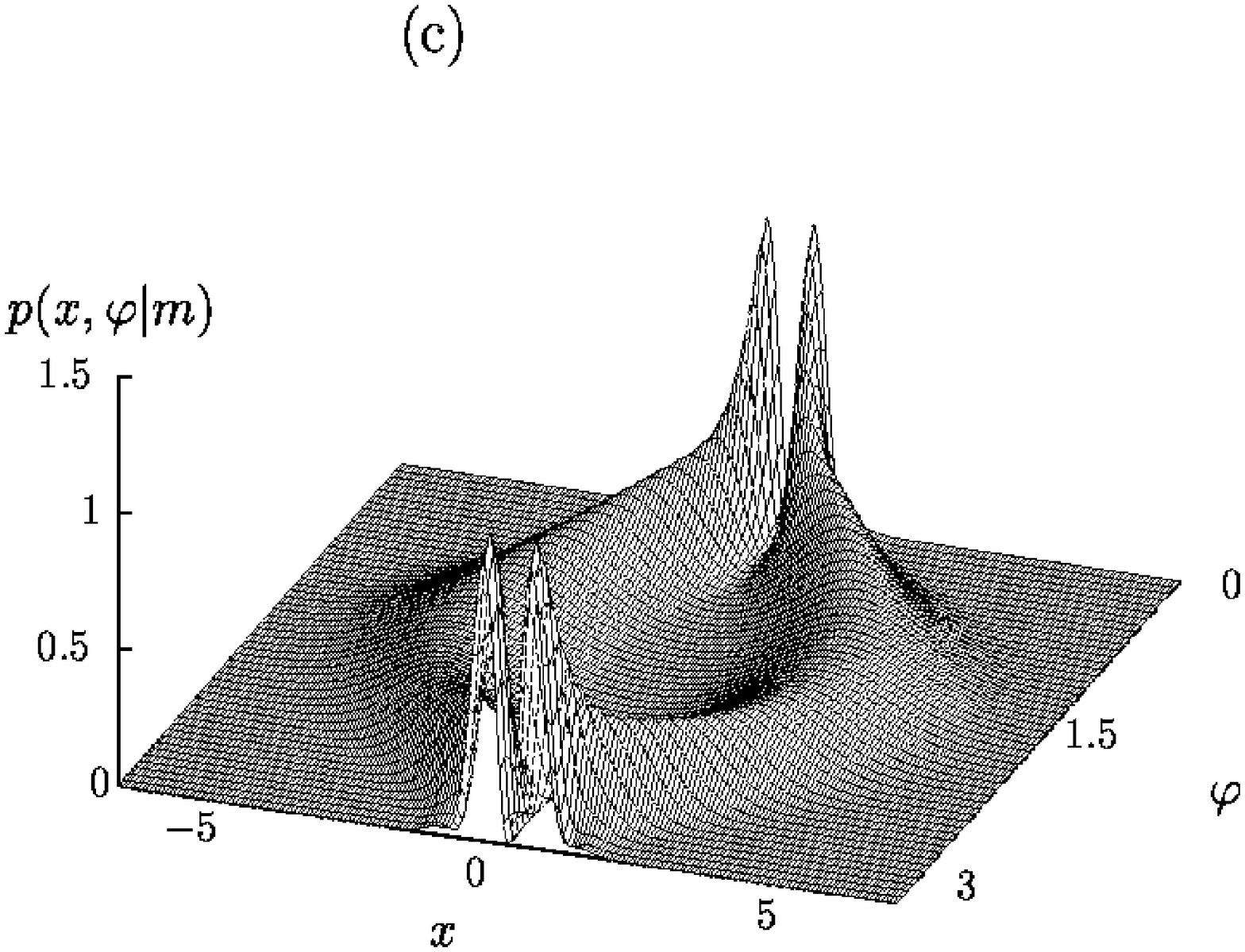,width=\linewidth}
\end{minipage}\hfill
\begin{minipage}[b]{0.5\linewidth}
\centering\epsfig{figure=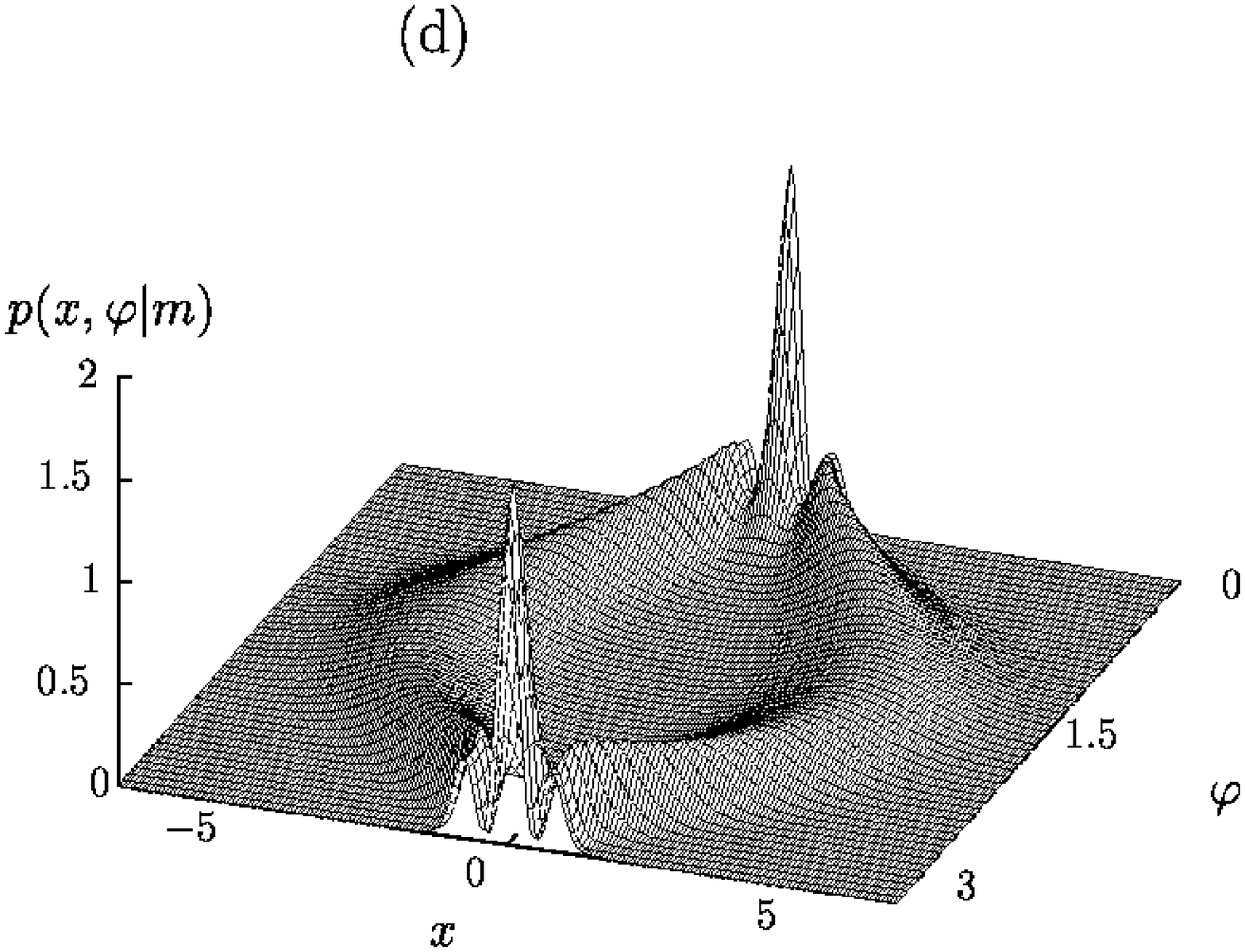,width=\linewidth}
\end{minipage}
\caption{ \label{fig1}Quadrature distribution $p(x,\varphi|m)$
of the conditional state $|\Psi_{m}\rangle$ for
$\alpha \!=\! 0.6$ and various numbers $m$ of measured photons
[(a) $m\!=\!1$, (b) $m\!=\!2$, (c) $m\!=\!3$, (d) $m\!=\!4$].
} 
\end{figure}

   From Fig.~\ref{fig1} we see that for $\varphi$ near $\pi/2$
the quadrature-component distribution $p(x,\varphi|m)$ ($m\!>\!0$)
exhibits two separated peaks, whereas for $\varphi$ close to $0$ or $\pi$
an interference pattern is observed. It should be noted that
although the quadrature-component distribution (\ref{F4})
bears a strong resemblance to that obtained in Ref.~\cite{Yurke2},
a more detailed comparison shows that they are different.
Clearly, the beam splitter transformation,
Eqs.~(\ref{1.03}) -- (\ref{1.03b}), can not be identified, in general,
with the transformation in the back-action-evading scheme considered in
Ref.~\cite{Yurke2}.

%%%%%%%%%%%%%%%%%%%%%%%%%%%%%%%%%%%%%%%%%%%%%%%%%%%%%%%%%%%

\subsection{Wigner function}
\label{sec31}

Using  Eqs.~(\ref{1.4}) and (\ref{F3}) together with the relations
(\ref{A.2}) and (\ref{A.3}), the Wigner function $W(x,p|m)$ of the state
$|\Psi_{m}\rangle$,
\begin{eqnarray}
\label{W1}
W(x,p|m) = \frac{1}{\pi}\int\limits_{-\infty}^{+\infty} {d}y \, e^{2ipy}
\langle x\!-\!y|\Psi_{m}\rangle\langle\Psi_{m}|x\!+\!y\rangle,
\end{eqnarray}
can be calculated in a straightforward way. We obtain
\begin{eqnarray}
\lefteqn{
W(x,p|m)= 
\frac{2\alpha^{m} e^{-\lambda x^2}}{\pi^{3/2}
{\cal N}_{m} [2(\alpha+1)]^{m+1}}
\int\limits_{-\infty}^{+\infty} {d}y
\Bigg\{
e^{-\lambda y^2 + 2i p y}
}
\nonumber \\ && \hspace{2ex} \times \,
{\rm H}_{m}\!\left[i\sqrt{\frac{\alpha}{1\!+\!\alpha}}(y\!-\!x)\right]
{\rm H}_{m}\!\left[i\sqrt{\frac{\alpha}{1\!+\!\alpha}}(x\!+\!y)\right]
\Bigg\},
\label{W2}
\end{eqnarray}
\begin{equation}
\lambda = \frac{1-\alpha}{1+\alpha} \,,
\label{W3}
\end{equation}
which after calculation of the integral \cite{Gradshteyn1}
can be given by
\begin{eqnarray}
\label{W4}
\lefteqn{
W(x,p|m)=
\frac{1}{\pi{\cal N}_{1m}}
\exp\!\left( - \lambda x^2 - \frac{p^2}{\lambda} \right)
}
\nonumber \\ && \hspace{2ex} \times
\sum_{k=0}^{m}
\frac{(-2|\alpha |)^{k}}{k! [ (m-k)! ]^{2}}
\left |{\rm H}_{m-k}\!\left[i\sqrt{\alpha\lambda}\;
\left(x+i\frac{p}{\lambda}\right)\right]\right |^2,
\end{eqnarray}
where
\begin{equation}
{\cal N}_{1m} =
\sum_{k=0}^{\left[ \frac{1}{2} m \right] }
\frac{(2 |\alpha |)^{m-2k}}{(m-2k)! \left( k! \right) ^{2}} \, .
\end{equation}

\begin{figure}
\noindent
\put(90,180){\large{(a)}}
\put(300,180){\large{(b)}}
\begin{minipage}[b]{0.5\linewidth}
\centering\epsfig{figure=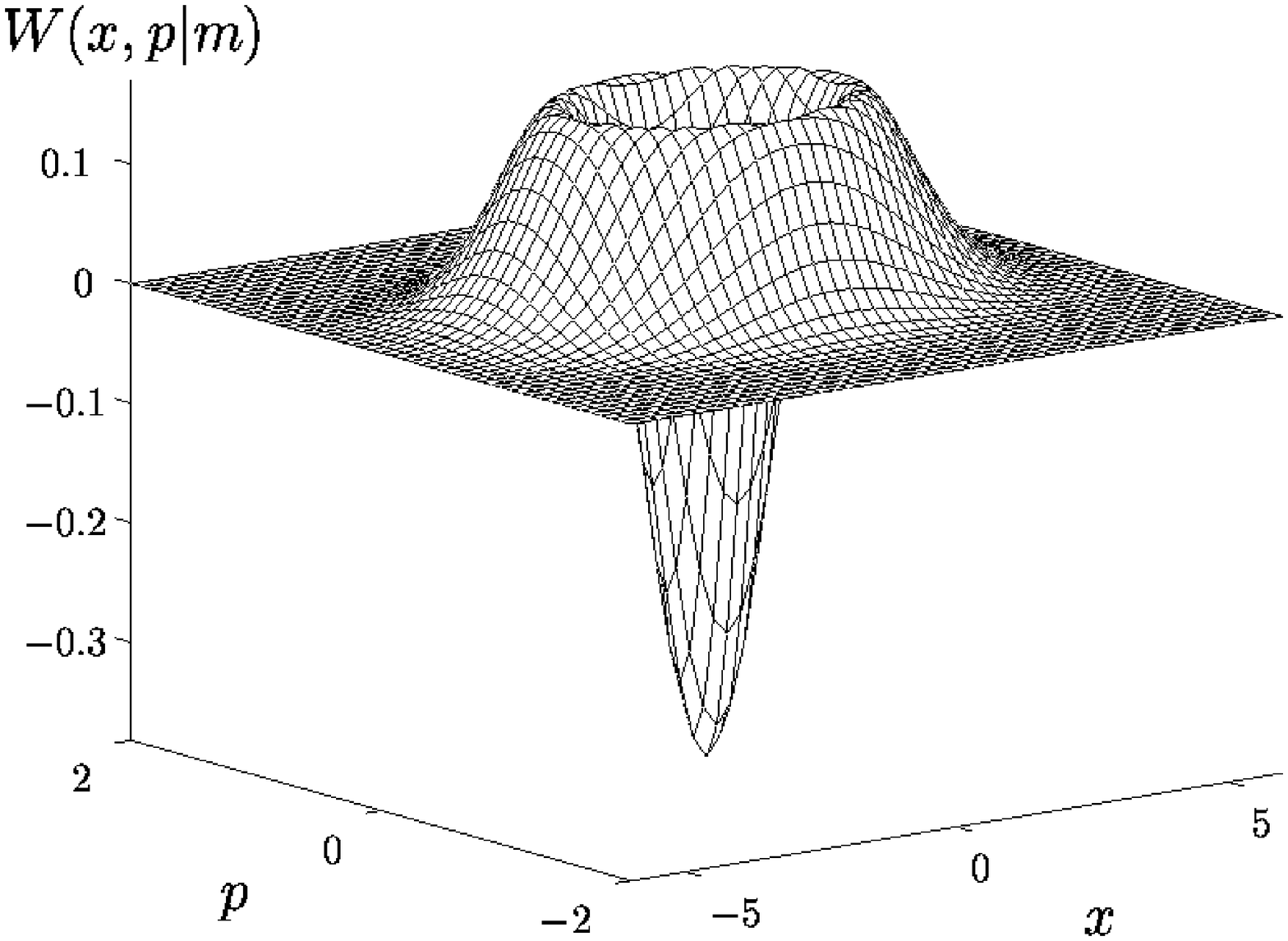,width=\linewidth}
\end{minipage}\hfill
\begin{minipage}[b]{0.5\linewidth}
\centering\epsfig{figure=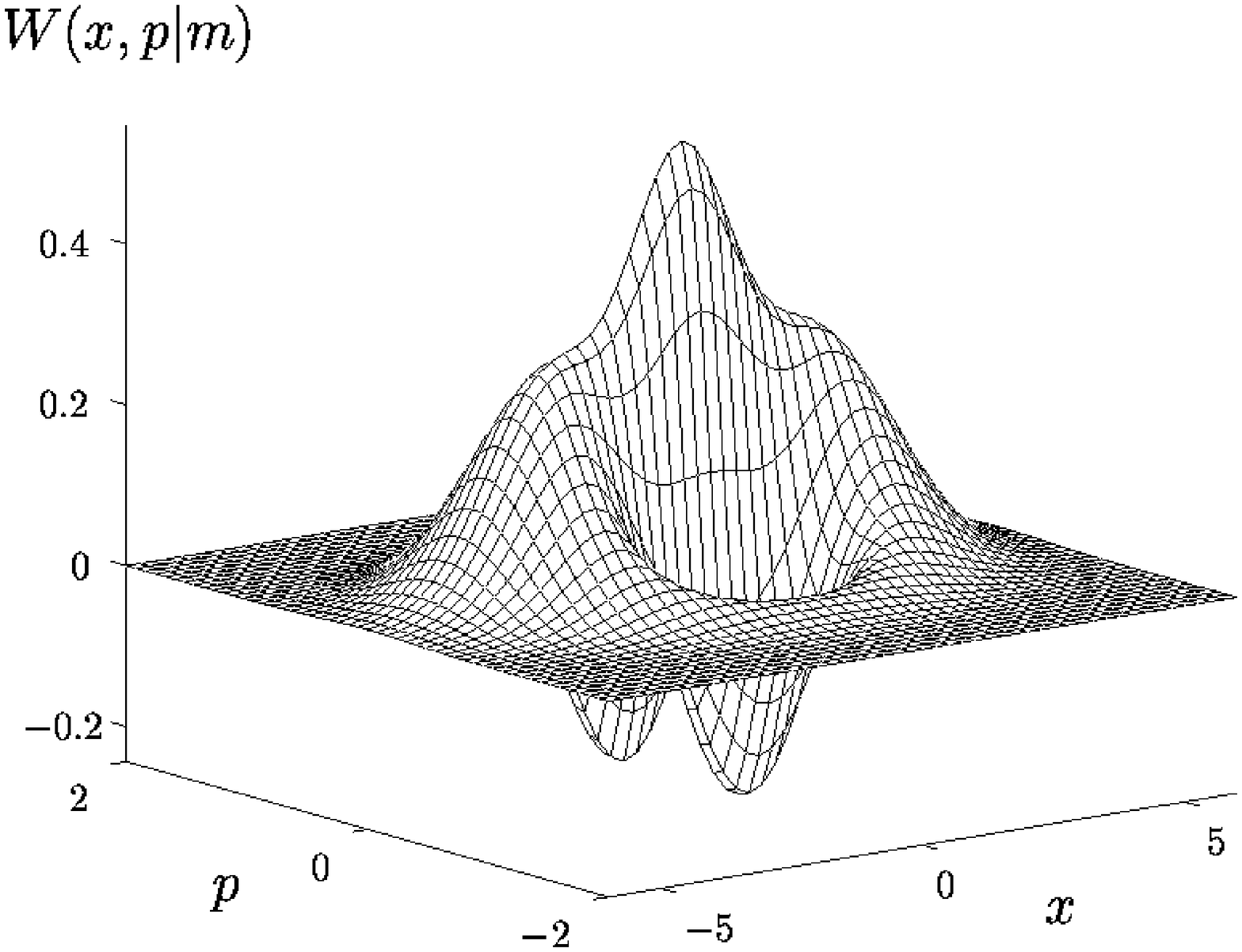,width=\linewidth}
\end{minipage}

~

\vspace{2cm}

~
\put(90,180){\large{(c)}}
\put(300,180){\large{(d)}}
\begin{minipage}[b]{0.5\linewidth}
\centering\epsfig{figure=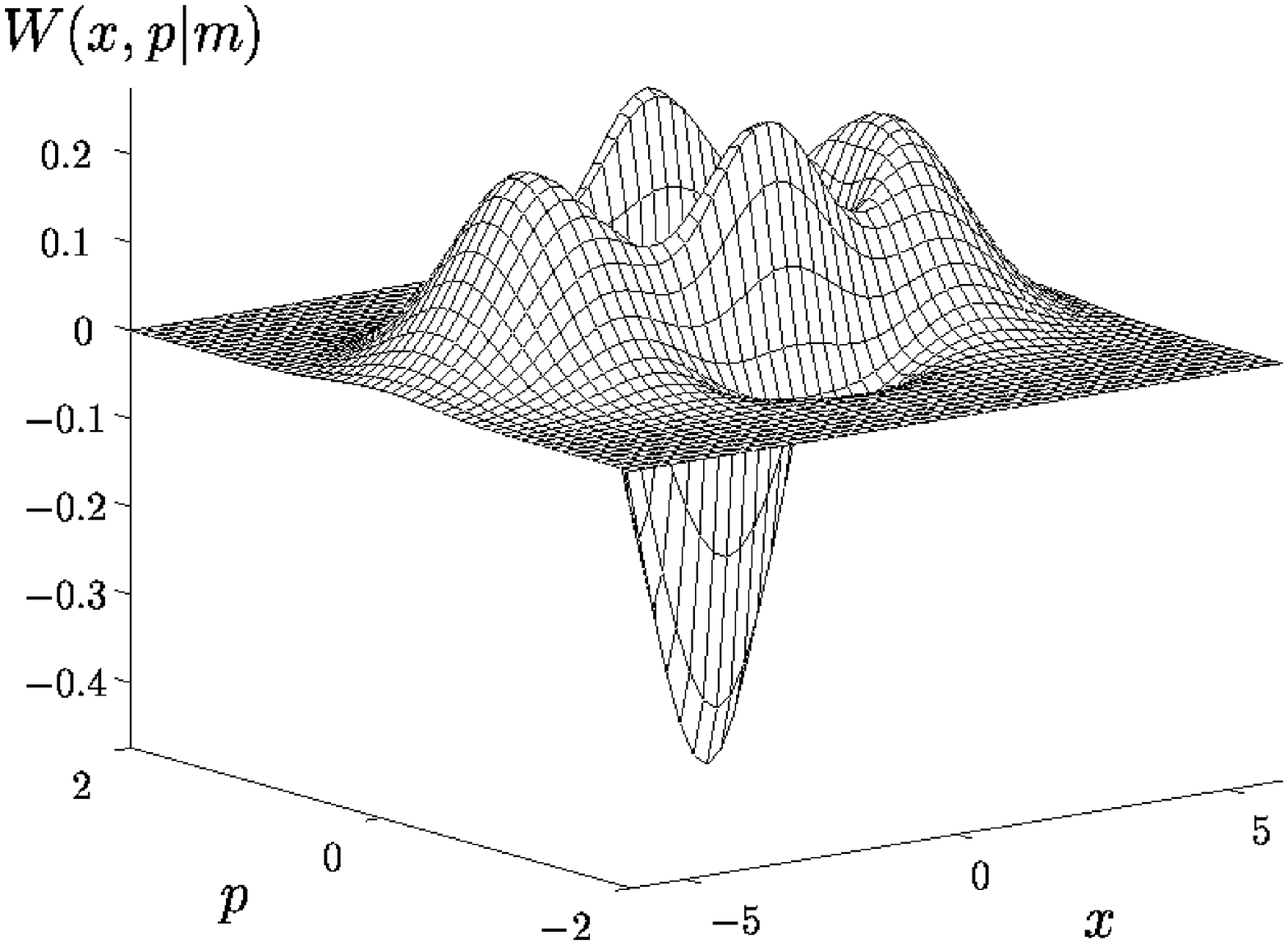,width=\linewidth}
\end{minipage}\hfill
\begin{minipage}[b]{0.5\linewidth}
\centering\epsfig{figure=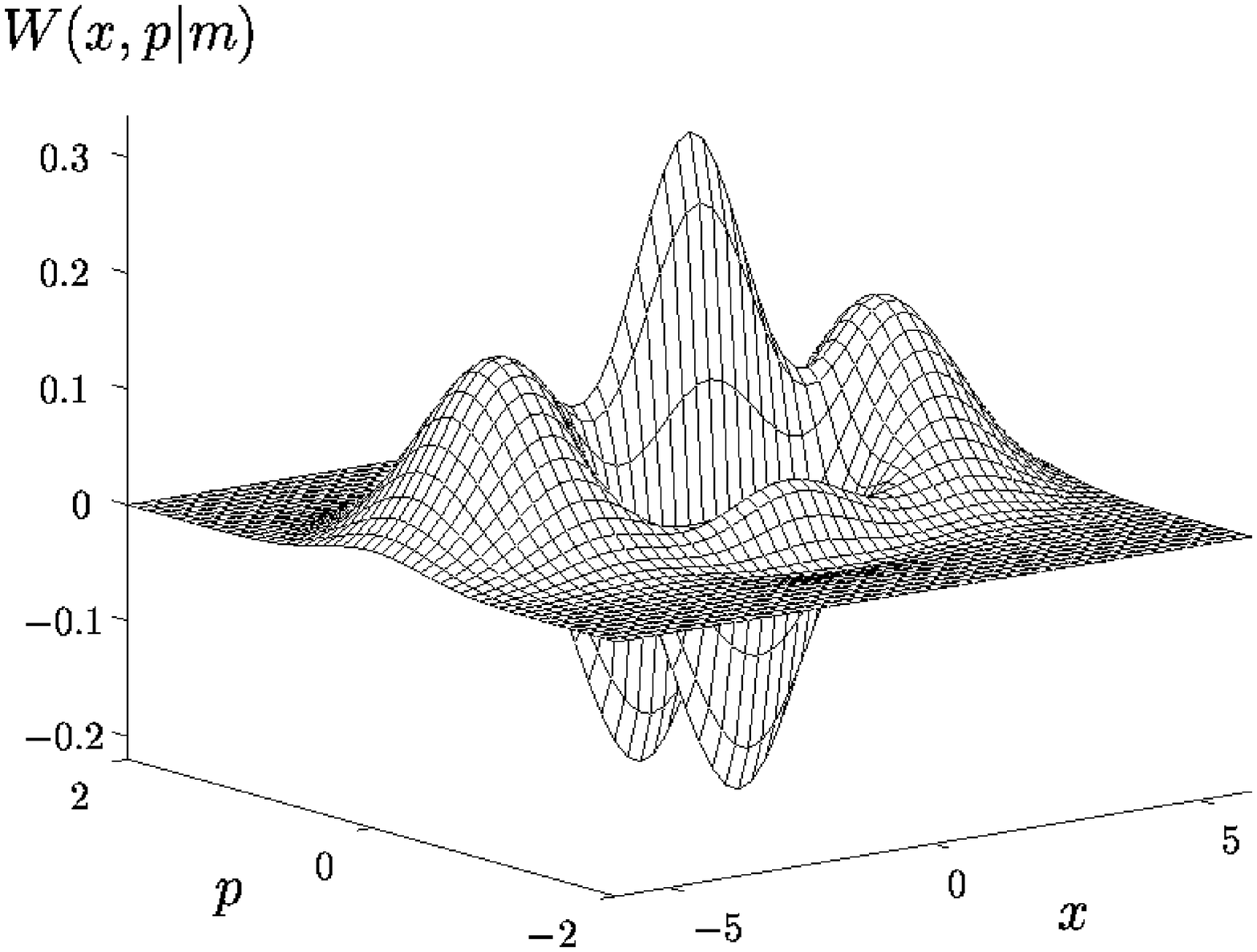,width=\linewidth}
\end{minipage}
\caption{ \label{fig2}Wigner function $W(x,p|m)$
of the conditional state $|\Psi_{m}\rangle$ for
$\alpha \!=\! 0.6$ and various numbers $m$ of measured photons 
[(a) $m\!=\!1$, (b) $m\!=\!2$, (c) $m\!=\!3$, (d) $m\!=\!4$].
} 
\end{figure}

Equation~(\ref{W4}) reveals that when no photons are detected,
$m\!=\!0$, then the Wigner function exhibits a single peak and
simply corresponds to a squeezed-vacuum Gaussian, as already
mentioned in Sec.~\ref{sec2}. Examples of $W(x,p|m)$ ($m\!>\!0$)
are shown in Fig.~\ref{fig2}. From the plots two separated peaks
and an oscillatory regime between them can be seen -- a shape that is
typical of Schr\"{o}dinger cat-like states. The separation of the two
peaks is seen to increase with the number $m$ of detected photons.
Since with increasing $m$ the number of photons in the conditional state
$|\Psi_{m}\rangle$ also increases, which is a consequence of the beam
splitter transformation, the behavior is quite similar to that of a
superposition of two coherent states.

%%%%%%%%%%%%%%%%%%%%%%%%%%%%%%%%%%%%%%%%%%%%%%%%%%%%%%%%%%%%%%%%%%%%

\subsection{Husimi function}
\label{sec32}

The Husimi function $Q(x,p|m)$ of the state $|\Psi_{m}\rangle$ is
defined by
\begin{equation}
Q(x,p|m)= \frac{1}{2\pi} \, |\langle\beta|\Psi_m\rangle|^2 ,
\label{Q1}
\end{equation}
where $|\beta \rangle $ is a coherent state, and
$\beta=2^{-1/2}(x+ip)$. It is worth noting that, contrary
to the Wigner function, the Husimi function is a phase-space
function that can directly be measured in multiport balanced homodyning
using six-port \cite{Walker1,Zucchetti1} or eight-port schemes
\cite{Walker2,Freyberger1}.
Expanding $|\beta\rangle$ in the Fock basis,
\begin{equation}
|\beta\rangle = e^{-|\beta|^2/2}\sum_{n=0}^{\infty}\frac{\beta^n}
{\sqrt{n!}}\,|n\rangle,
\label{Q2}
\end{equation}
and recalling Eq.~(\ref{1.4}) and the relations (\ref{A.2}) and (\ref{A.4}),
the scalar product $\langle\beta|\Psi_m\rangle $ can easily be calculated.
After some algebra we find that $Q(x,p|m)$ can be written as
\begin{eqnarray}
\label{Q3}
\lefteqn{
Q(x,p|m)=\frac{|\alpha |^m}{\pi{\cal N}_{m}2^{m+1}}
\left |{\rm H}_{m}\!\left[\textstyle\frac{1}{2} i \sqrt{\alpha}
(x+ip) \right] \right |^2
}
\nonumber \\ && \hspace{12ex} \times \,
\exp\!\left\{ -\textstyle\frac{1}{2}
\left[ (1-\alpha)x^2+(1+\alpha)p^2 \right] \right\}.
\end{eqnarray}
As expected, for $m\!=\!0$ the Husimi function is Gaussian, whereas
for $m\!>\!0$ a two peak-structure is observed. Since the
Husimi function is always non-negative, the interference properties
of the state are not so apparent as in the case of the Wigner function.

%%%%%%%%%%%%%%%%%%%%%%%%%%%%%%%%%%%%%%%%%%%%%%%%%%%%%%%%%%%%%%%%%%%%%%

\subsection{Component states}
\label{sec35}

Schr\"{o}dinger cat-like states are commonly defined as
superpositions of two macroscopically distinguishable quantum states. From
Eqs.~(\ref{1.4}) and (\ref{1.5}) it is seen that $|\Psi_{m}\rangle$
can be regarded as a superposition of states $|\Psi_{m}^{(+)}\rangle$
and $|\Psi_{m}^{(-)}\rangle$ as follows:
\begin{eqnarray}
\label{Co1}
|\Psi_m\rangle = A\,
\left(|\Psi^{(\!+\!)}_m\rangle +|\Psi ^{(\!-\!)}_m\rangle\right) ,
\end{eqnarray}
where 
\begin{equation}
|\Psi ^{(\pm)}_m\rangle=\frac{1}{\sqrt{{\cal N}_m^{(\pm)}}}
\sum_{n=0}^{\infty}
c^{(\pm)}_{m,n}(\alpha) |n\rangle ,
\label{Co2}
\end{equation}
with
\begin{equation}
c^{(\pm)}_{m,n}(\alpha)=\frac{(n+m)!}
{\Gamma[ (n+m)/2 + 1 ]
\sqrt{n!}}
\left(\pm\sqrt{\textstyle\frac{1}{2}\alpha}\right)^{n+m}\!\!.
\label{Co3}
\end{equation}
The normalization factor ${\cal N}_{m}^{(\pm)}$ is calculated to be
[see Eqs.~(\ref{B3}) and (\ref{B11})]
\begin{eqnarray}
\lefteqn{
{\cal N}_m^{(\pm)}
=2 {\cal N}_m
- \frac{m!^2(-|\alpha|)^m}{\sqrt{\pi}2^m\Gamma(m+3/2)}
}
\nonumber \\ && \hspace{5ex} \times \,
{\rm F}\!\left[
\textstyle\frac{1}{2}(m\!+\!1),\textstyle\frac{1}{2}(m\!+\!1),
\textstyle\frac{1}{2}(2m\!+\!3),1-\alpha^2 \right] ,
\label{CO4}
\end{eqnarray}
${\cal N}_{m}$ being given in Eq.~(\ref{1.3b}). In Eq.~(\ref{CO4})
${\rm F}(\alpha,\beta,\gamma,z)$ is the hypergeometric function,
and the normalization constant $A$ in Eq.~(\ref{Co1}) is simply given
by $A\!=\!\frac{1}{2}({\cal N}_m^{(\pm)}/{\cal N}_m)^{1/2}$.
Note that ${\cal N}_m^{(\pm)}\!\approx\! 2{\cal N}_m$, i.e.,
$A\!\approx\!1/\sqrt{2}$ for larger $m$ (the approximation is
very good for $m\!>\!4$).

\begin{figure}
\centering\epsfig{figure=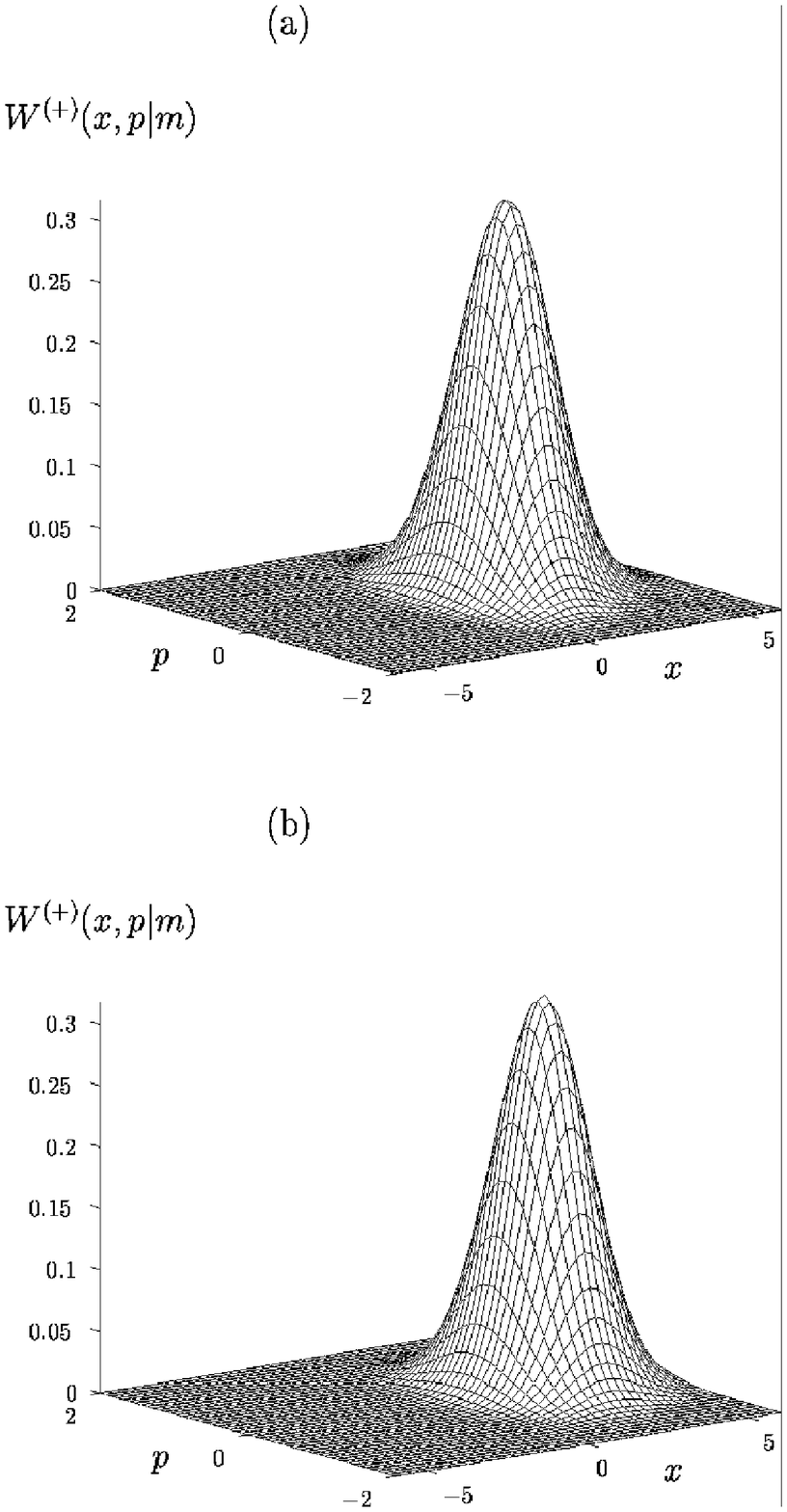,width=1\linewidth}
\caption{ \label{fig4}Wigner function $W^{(+)}(x,p|m)$
of the state $|\Psi^{(\!+\!)}_m\rangle$ for
$\alpha\! =\! 0.6$ and various numbers $m$ of measured photons 
[(a) $m\!=\!1$, (b) $m\!=\!3$].
  }
\end{figure}

Plots of the Wigner function of the states $|\Psi^{(+)}_{m}\rangle$
are given in Fig.~\ref{fig4}. The behavior of the states
$|\Psi^{(-)}_{m}\rangle$ is quite similar. From the figure we see
that the states $|\Psi ^{(+)}_m\rangle$ (and also the states
$|\Psi ^{(-)}_m\rangle$) are very close to squeezed coherent states.
For chosen $\alpha$ the squeezing effect decreases with increasing $m$.
The small deviation from Gaussian states is indicated by the negative values
of the Wigner function (which in our case are of the order of magnitude
of $-10^{-5}$). To illustrate the difference between the states
$|\Psi^{(\pm)}_{m}\rangle$ and the squeezed coherent states, let us consider
the Husimi function $Q^{(\pm)}(x,p|m)$ of $|\Psi^{(\pm)}_{m}\rangle$.
According to Eq.~(\ref{B13}) we obtain
\begin{eqnarray}
\lefteqn{
Q^{(\pm)}(x,p|m) =
\frac{\alpha^m(m!)^2}{\pi^2{\cal N}_m^{(\pm)}} \, e^{-|\beta|^2}
}
\nonumber \\ && \hspace{5ex} \times \,
\exp\!\left[\textstyle\frac{1}{4}\alpha(\beta^2+{\beta^\ast}^2)\right]
\left| {\rm D}_{-m-1}(\pm\sqrt{\alpha}\beta) \right|^2,
\label{CO5}
\end{eqnarray}
where $\beta = (x+ip)/\sqrt{2}$, and ${\rm D}_{m}(z)$ is
the parabolic cylinder function (which indicates the deviation from
a Gaussian). Recalling the asymptotic behavior of the ${\rm D}_{-m-1}(z)$
for large values of $m$ \cite{Erdelyi1}, we find that
\begin{eqnarray}
\lefteqn{
Q^{(\pm)}(x,p|m)
}
\nonumber \\ && \hspace{5ex}
\propto\,
\exp\!\left[-|\beta|^2\pm\sqrt{\alpha m}(\beta+\beta^\ast)-
|\sqrt{\alpha m}|^2\right]
\label{Co6}
\end{eqnarray}
($ m\!\to\!\infty$), which corresponds to the scalar product
$|\langle\beta|\gamma\rangle|^2$ between the coherent states
$|\beta\rangle$ and $|\gamma\rangle$, with $\gamma\!=\!\pm\sqrt{\alpha m}$.
Thus for sufficiently large numbers of detected photons, $m$, the component
states $|\Psi_{m}^{(\pm)}\rangle$ approach coherent
states. This kind of behavior has also been found in Ref.~\cite{Yurke2}
for the states studied therein.

%%%%%%%%%%%%%%%%%%%%%%%%%%%%%%%%%%%%%%%%%%%%%%%%%%%%%%%%%%%

\section{Realistic photon counting}
\label{sec4}

In order to produce the conditional states $|\Psi_{m}\rangle$, highly
efficient and precise photocounting is needed. Unfortunately, there have
been no highly efficient photodetectors available which precisely
distinguish between $m$ and $m\!+\!1$ photons. Recently the
proposal has been made to measure the photon-number statistics
using photon chopping \cite{Paul1}. The mode to be detected is used as
an input of an optical $2N$-port, the other input ports being unused.
To each of the output ports a highly efficient avalanche photodiode is
connected which can distinguish between photons being present or absent.
The photon-number statistics of the input mode can then be obtained from
the recorded output coincident-event statistics.

In particular, if $m$ photons are present in the input, the probability
of recording $k$ coincident events is given by \cite{Paul1}
\begin{equation}
  \label{T1}
  \tilde P_{N}(k|m) = \frac{1}{N^{m}} {N \choose k}
  \sum_{l=0}^{k}(-1)^{l} {k \choose l} (k - l)^{m}
\label{4.1}
\end{equation}
for $k \le m$, and $\tilde P_{N}(k|m)\!=\!0$ for $k\!>\!m$. Note that
$\tilde P_{N}(k|m) \!\to\! \delta_{k,m}$ for $N \!\to\! \infty$.
In Eq.~(\ref{4.1}) perfect detection is assumed. The
effect of nonperfect detection corresponds to a random process such that
photons are excluded from detection with probability $1\!-\!\eta$,
$\eta$ being the efficiency of the photodiodes ($0<\eta\!\leq\!1$).
The probability of recording $k$ coincident events then modifies to
\begin{equation}
  \label{T2}
  \tilde P_{N,\eta}(k|m) = \sum_{l} \tilde P_{N}(k|l) \, M_{l,m}(\eta) ,
\end{equation}
where the matrix $M_{l,m}(\eta)$ is given by
\begin{equation}
  \label{eq:Pmatrix}
  {M}_{l,m}(\eta) = {m \choose l} \eta^{l} (1 - \eta )^{m-l}
\end{equation}
for $l\!\le\!m$, and ${M}_{l,m}(\eta)\!=\!0$ for $l\!>\!m$.

Since detection of $k$ coincident events can result from various numbers
$m$ of photons, the conditional state is in general a statistical mixture.
Therefore in place of Eq.~(\ref{1.3}) we now have
\begin{eqnarray}
\label{T3}
\hat \varrho_{\rm out}(k)
= \sum_{m}  P_{N,\eta}(m|k) \,|\Psi_{m} \rangle\langle \Psi_{m} |,
\end{eqnarray}
where $|\Psi_{m}\rangle$ is given in Eq.~(\ref{1.4}), and
$P_{N,\eta}(m|k)$ is the probability of $m$ photons being present
under the condition that $k$ coincidences are recorded. The conditional
probability $P_{N,\eta}(m|k)$ can be obtained using the Bayes rule as
\begin{eqnarray}
\label{T4}
P_{N,\eta}(m|k) = \frac{1}{\tilde P_{N,\eta}(k)} \tilde P_{N,\eta}(k|m)
\ P(m).
\end{eqnarray}
Here $P(m)$ is the prior probability (\ref{P2}) of
$m$ photons being present, and accordingly, $\tilde P_{N,\eta}(k)$
is the prior probability of recording $k$ coincident events,
\begin{eqnarray}
\label{T5}
\tilde P_{N,\eta}(k) = \sum_{m} \tilde P_{N,\eta}(k|m)
\ P(m) .
\end{eqnarray}

\begin{figure}
\noindent
\put(90,180){\large{(a)}}
\put(300,180){\large{(b)}}
\begin{minipage}[b]{0.5\linewidth}
\centering\epsfig{figure=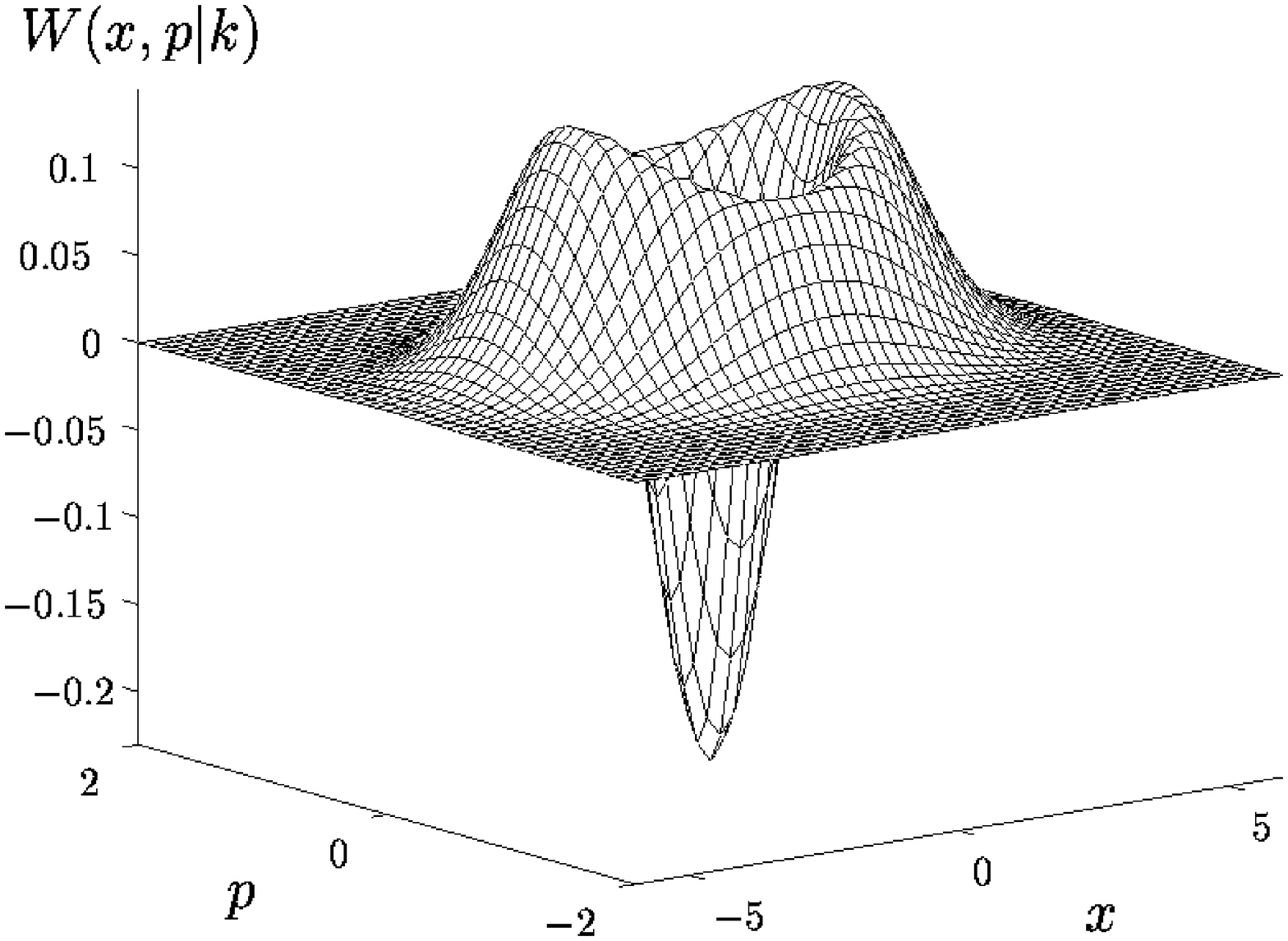,width=\linewidth}
\end{minipage}\hfill
\begin{minipage}[b]{0.5\linewidth}
\centering\epsfig{figure=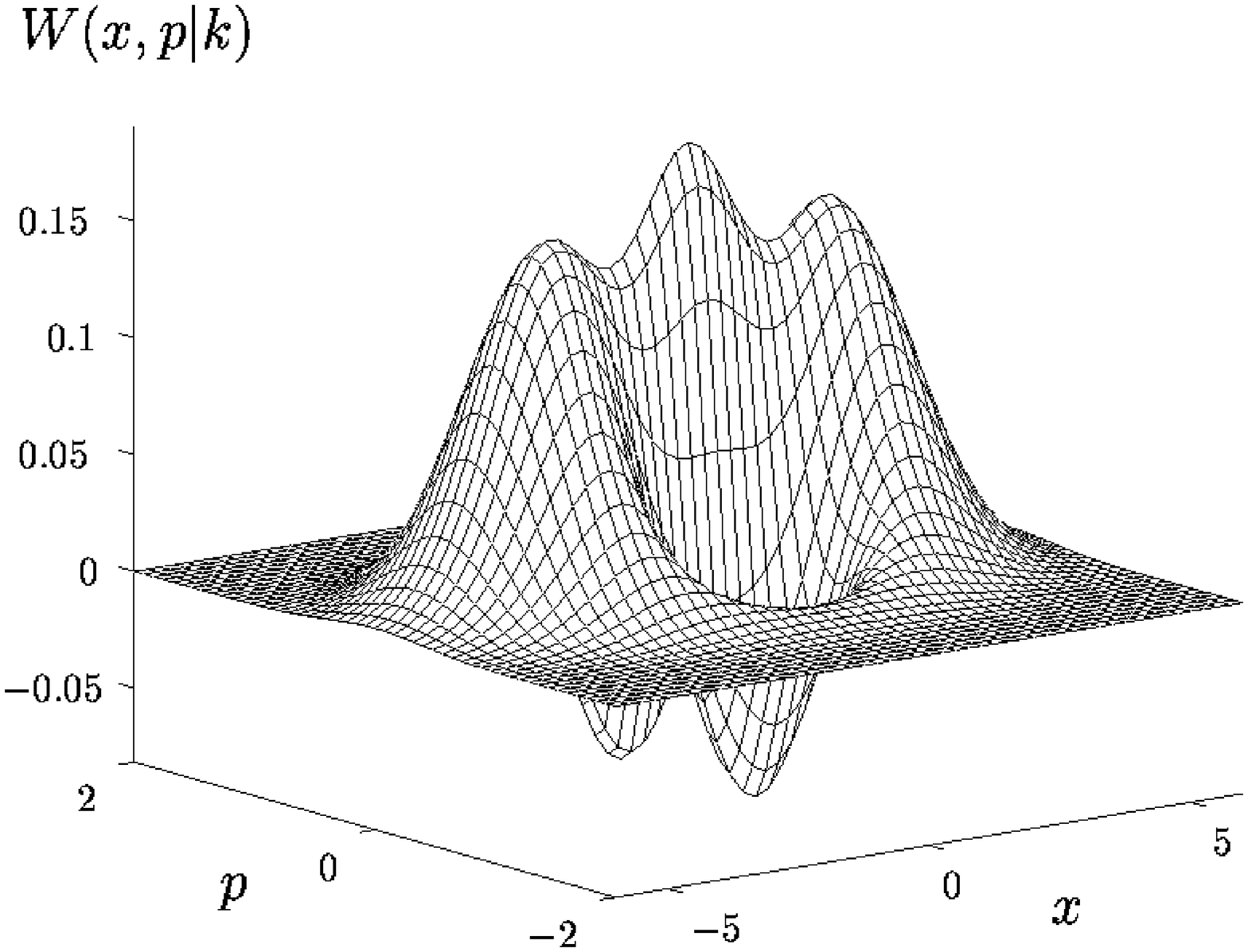,width=\linewidth}
\end{minipage}

~

\vspace{2cm}

~
\put(90,180){\large{(c)}}
\put(300,180){\large{(d)}}
\begin{minipage}[b]{0.5\linewidth}
\centering\epsfig{figure=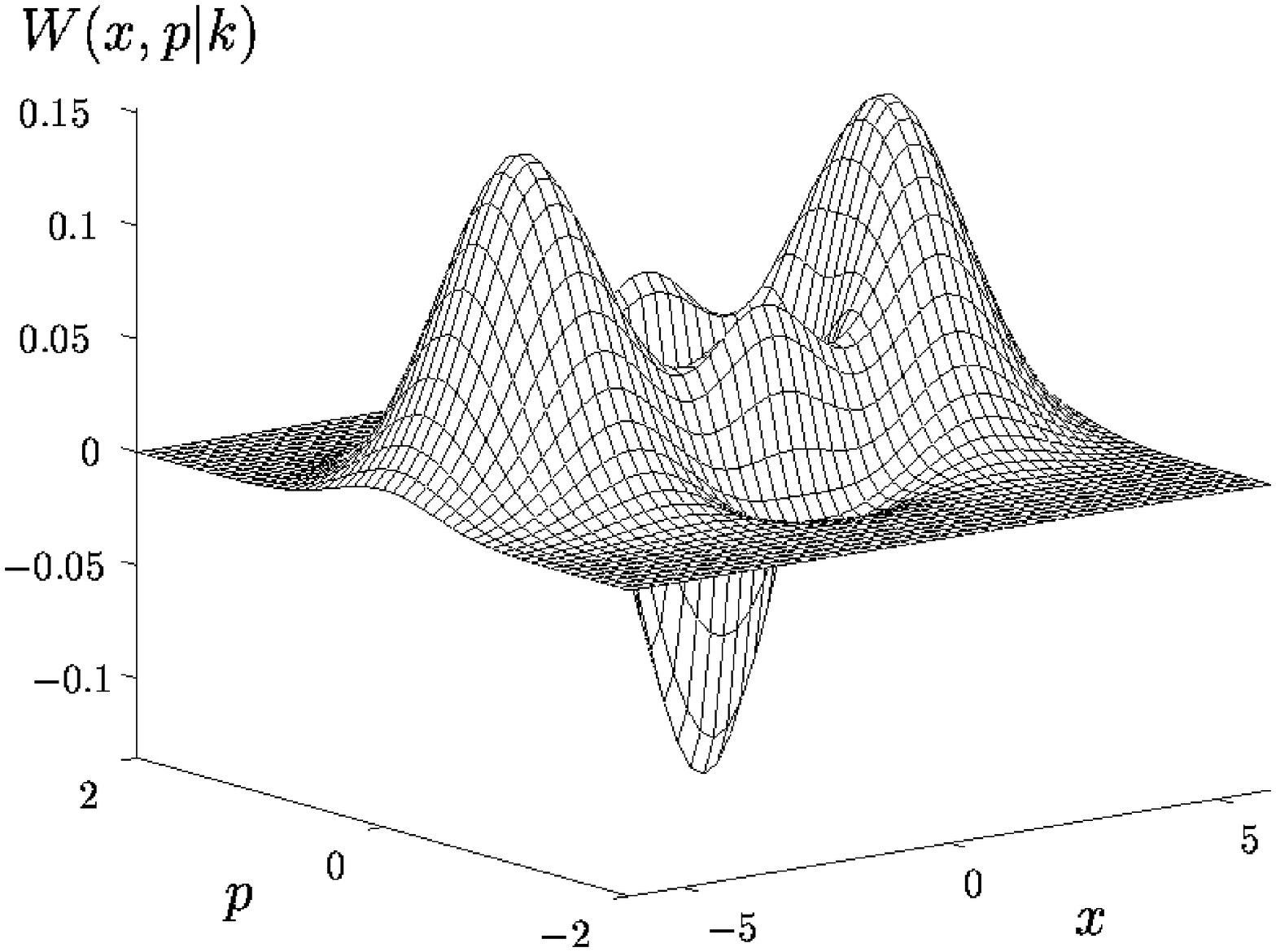,width=\linewidth}
\end{minipage}\hfill
\begin{minipage}[b]{0.5\linewidth}
\centering\epsfig{figure=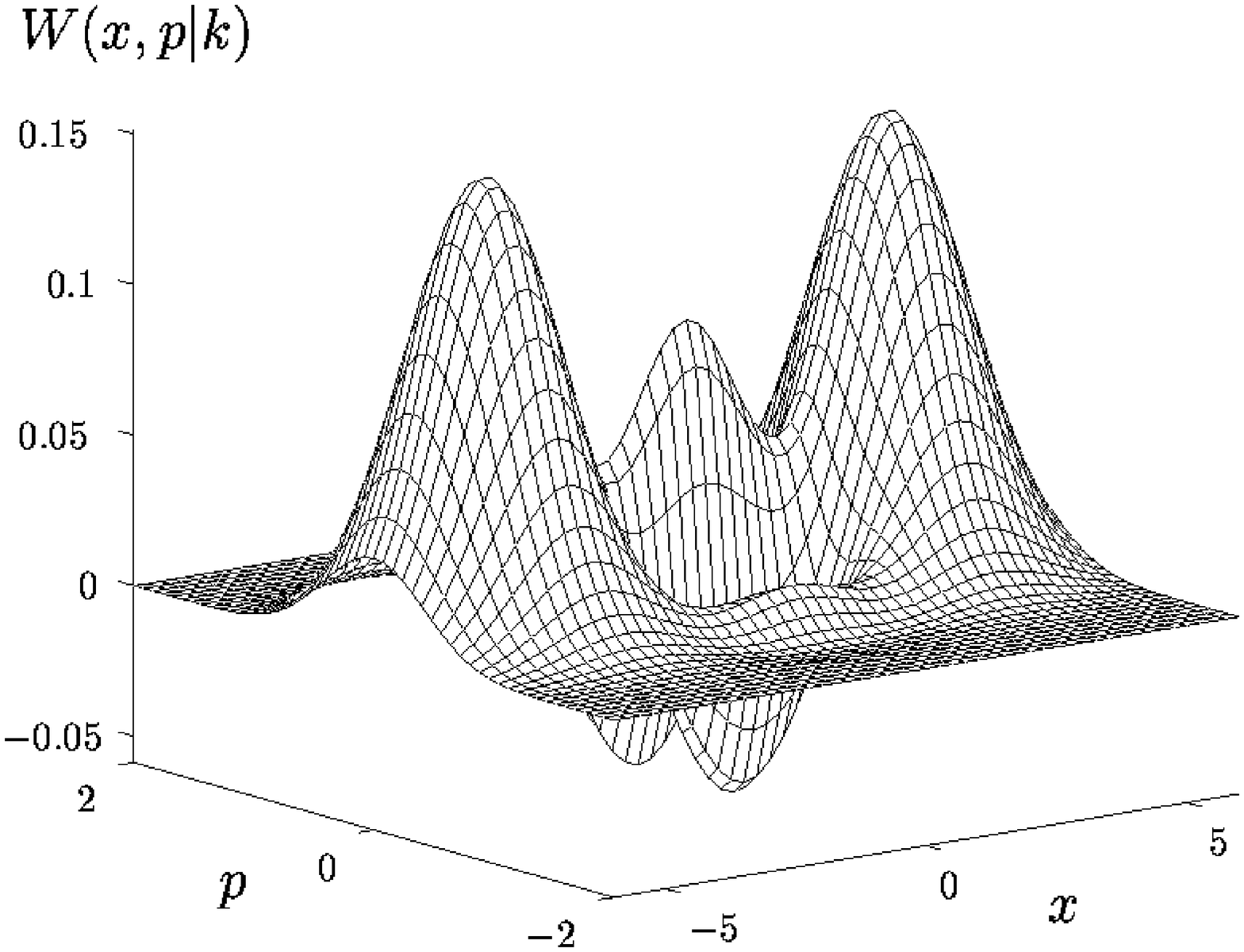,width=\linewidth}
\end{minipage}
\caption{ \label{fig5}Wigner function of the conditional state
with realistic photodetection ($N\!=\!10$, $\eta\! = \!0.8$)
for $\kappa = 0.75$, $|T|^2 = 0.8$
($\alpha\!=\!0.6$) and various numbers $k$ of coincident events
[(a) $k\!=\!1$, $\tilde P_{N,\eta}(k)=10.99\%$;
 (b) $k\!=\!2$, $\tilde P_{N,\eta}(k)=2.95\%$;
(c) $k\!=\!3$, $\tilde P_{N,\eta}(k)=0.69\%$;
(d) $k\!=\!4$, $\tilde P_{N,\eta}(k)=0.16\%$],
.the probabilities $\tilde P_{N,\eta}(k)$ of the coincidences being 
calculated according to 
Eq.~(\protect\ref{T5})
} 
\end{figure}

\begin{figure}
\centering\epsfig{figure=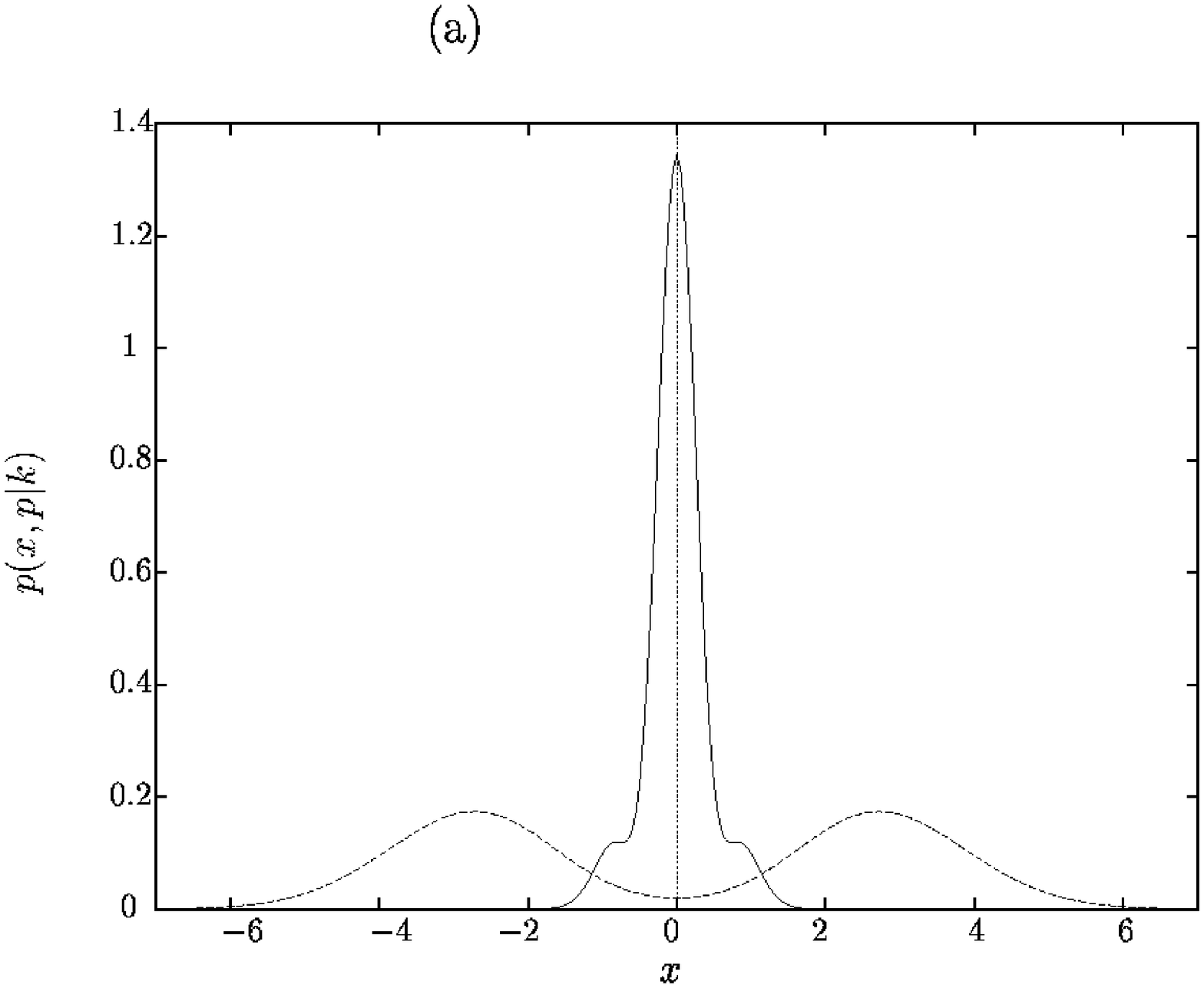,width=0.8\linewidth}
\centering\epsfig{figure=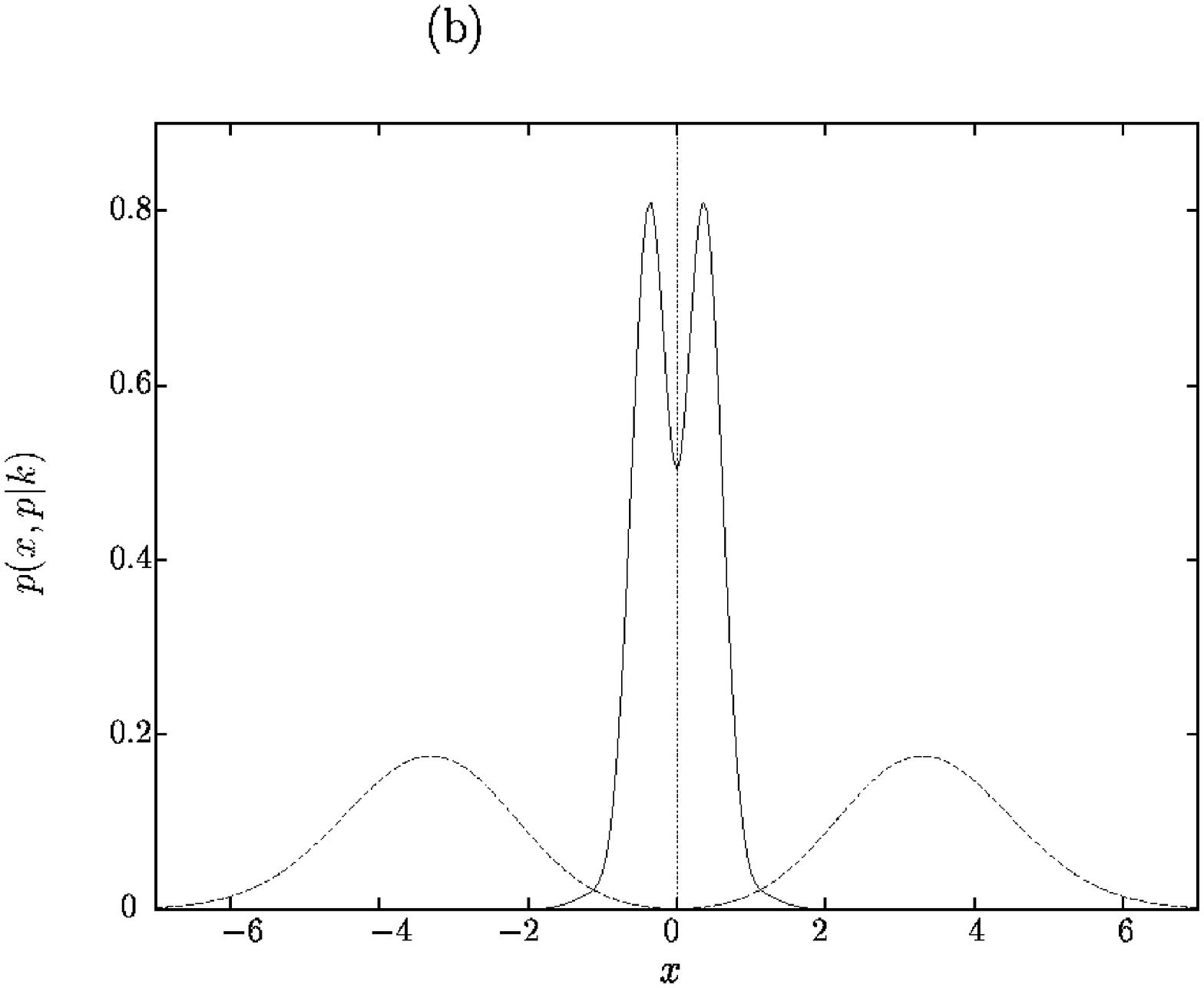,width=0.8\linewidth}
\caption{ \label{fig6}Quadrature distribution of the conditional state
with realistic photodetection ($N\!=\!10$, $\eta\! = \!0.8$)
for the phase parameters $\varphi \! = \! 0$ (full line) and
$\varphi \! = \! \pi /2$ (broken line), various numbers $k$
of coincident events [(a) $k\!=\!2$, (b) $k\!=\!3$], and $\kappa = 0.75$,
$|T|^2 = 0.8$ ($\alpha = 0.6$).}
\end{figure}

Examples of the Wigner function and quadrature-component distributions
of the mixed state (\ref{T3}) are plotted in Figs.~\ref{fig5} and
\ref{fig6}, respectively, We see that the quantum interferences can
still be preserved also for realistic values of the number of photodiodes
and their efficiencies, such as $N\!=\!10$ and $\eta\!=\,$80\%.
The interference structure is (for chosen $N$ and $\eta$) more and more
smeared with increasing $k$. Clearly, a larger number of detected
coincidences implies (for chosen $N$ and $\eta$) a larger probability
of ``losing'' some of the photons. Any lost photon switches
the parity of the conditional state (from even to odd and vice versa)
and therefore destroys the interference pattern.

%%%%%%%%%%%%%%%%%%%%%%%%%%%%%%%%%%%%%%%%%%%%%%%%%%%%%%%%%%%%%%%%%%%

\section{Conclusion}
\label{sec5}

We have shown that Schr\"{o}dinger cat-like states can be
generated by conditional measurements using a simple beam splitter
scheme. When a squeezed vacuum and an ordinary vacuum are mixed by a beam
splitter and in one of the output channels the number of photons
is measured, then the conditional quantum state in the other output
channel reveals all the properties of a Schr\"{o}dinger cat-like state.

To demonstrate this, we have analysed the conditional
states in terms of the photon-number and quadrature-component
distributions and the Wigner and Husimi functions and have
presented both analytical and numerical results. We have studied the
component states and shown that they are very close to squeezed
coherent states and approach coherent states for sufficiently large
numbers of detected photons. We have found that the basic features
of Schr\"{o}dinger cat-like states, such as the appearance of
two separated peaks and the interference pattern, become more pronounced
for larger values of the transmittance of the beam splitter,
the squeezing parameter of the input state, and the number of detected
photons. On the other hand, increasing transmittance implies a decrease of
the probability of photons being present, so that the ``better''
Schr\"{o}dinger cat-like states appear more rarely.

We have also discussed the problem of producing the
Schr\"{o}dinger cat-like states under the conditions of realistic
photocounting. For this purpose we have assumed multichannel detection
using highly efficient avalanche photodiodes. As expected, the measurement
smears the interference structure. However, for properly chosen parameters
the interference structure can still be found even for
a realistic arrangement of detectors.

\section*{Acknowledgment}
This work was supported by the Deutsche Forschungsgemeinschaft.

%%%%%%%%%%%%%%%%%%%%%%%%%%%%%%%%%%%%%%%%%%%%%%%%%%%%%%%%%%%

\begin{appendix}
\renewcommand{\thesection}{Appendix \Alph{section}}
\section{Sum. Rules}
\label{appA}
\setcounter{equation}{0}
\renewcommand{\theequation}{\Alph{section} \arabic{equation}}
Using the expansion in the Fock basis of $|\Psi_{m}\rangle$,
Eq.~(\ref{1.4}), various photon-number summations must be
performed in the further calculations, which can advantageously
be done by means of Hermite polynomials. For this purpose we rewrite
the coefficients $c_{m_{2},n_{1}}(\alpha)$, Eq.~(\ref{1.5}), as
\begin{equation}
c_{m_{2},n_{1}}(\alpha)=
\frac{[2(n+k+\delta)]!}
{(n+k+\delta)!\sqrt{(2n+\delta)!}}
\left(\textstyle\frac{1}{2}\alpha\right)^{n+k+\delta},
\label{A.1}
\end{equation}
where $\delta\!=\!0$ for $n_1\!=\!2n$ and $m_2\!=\!2k$, and
$\delta\!=\!1$ if $n_{1}\!=\!2n\!+\!1$ and $m_{2}\!=\!2k\!+\!1$.
Note that $c_{m_{2},n_{1}}(\alpha)\!=\!0$ otherwise. Recalling
the relation
\begin{equation}
{\rm H}_{2n}(0)=(-1)^n(2n)!/n! \,,
\label{A.1a}
\end{equation} 
we see that
\begin{equation}
c_{m_{2},n_{1}}(\alpha)=\frac{1}{\sqrt{(2n+\delta)!}} \,
{\rm H}_{2(n+k+\delta)}(0)
\left(-\textstyle\frac{1}{2}\alpha\right)^{n+k+\delta} .
\label{A.2}
\end{equation}
This enables us to apply standard summation rules, such as
Mehler's formula \cite{Erdelyi1}
\begin{eqnarray}
\lefteqn{
\sum_{k=0}^{\infty}\frac{1}{k!}{\rm H}_{k}(x)
{\rm H}_{k}(y)\left(\textstyle\frac{1}{2}z\right)^k
}
\nonumber \\ && \hspace{5ex} \, \,
=\, \frac{1}{\sqrt{1-z^2}}
\exp\!\left[\frac{2xyz-(x^2+y^2)z^2}{1-z^2}\right].
\label{A.9}
\end{eqnarray}
Taking the $l$th and $j$th derivatives with respect to $x$ and $y$,
respectively, of both sides of this equation and using the relation
\begin{eqnarray}
\frac{d^j}{dx^j}{\rm H}_{k}(x)=2^j\frac{k!}{(k-j)!}{\rm H}_{k-j}(x),
\label{A.2b}
\end{eqnarray}
we derive that
\begin{eqnarray}
\lefteqn{
\sum_{k=0}^{\infty}\frac{1}{k!}{\rm H}_{k+l}(x) {\rm H}_{k+j}(y)
\left(\textstyle\frac{1}{2}z\right)^k=\frac{1}{(1-z^2)^{(l+j+1)/2}}
}
\nonumber \\ && \hspace{2ex} \times \,
\exp\!\left[\frac{2xyz-(x^2+y^2)z^2}{1-z^2}\right]
\sum_{k=0}^{\min(l,j)}{l\choose k}{j\choose k}
 (2z)^kk!
\nonumber \\ && \hspace{2ex} \times \,
{\rm H}_{l-k}\!\left( \frac{x-zy}{\sqrt{1-z^2}}\right)
{\rm H}_{j-k}\!\left( \frac{y-zx}{\sqrt{1-z^2}}\right) .
\label{A.3}
\end{eqnarray}

Another useful sum rule is
\begin{eqnarray}
\lefteqn{
\sum_{k=0}^{\infty}\frac{1}{k!}{\rm H}_{k+j}(x)
\left(\textstyle\frac{1}{2}z\right)^k
}
\nonumber \\ && \hspace{5ex}
= \, \exp\!\left(xz-\textstyle\frac{1}{4}z^2\right)
{\rm H}_{k+j}\left(x-\textstyle\frac{1}{2}z\right),
\label{A.4}
\end{eqnarray}
which may be derived by taking the $j$th derivative of
the  generating function of the Hermite polynomials,
\begin{equation}
\sum_{k=0}^{\infty}{\rm H}_{k+j}(x)\frac{z^k}{k!}=
\exp\!\left(2xz-z^2\right) .
\label{A.4a}
\end{equation}

%%%%%%%%%%%%%%%%%%%%%%%%%%%%%%%%%%%%%%%%%%%%%%%%%%%%%
\section{Derivation of the relations (\ref{CO4}) and (\ref{CO5})}
\setcounter{equation}{0}
\label{appB}
From Eqs.~(\ref{1.4}), (\ref{1.5}) and (\ref{Co1}) -- (\ref{Co3})
we see that
\begin{eqnarray}
c_{m,n}(\alpha) =\textstyle\frac{1}{2}
\left[c^{(+)}_{m,n}(\alpha)+c^{(-)}_{m,n}(\alpha)\right] ,
\label{B1}
\end{eqnarray}
and hence
\begin{eqnarray}
\lefteqn{ 
\sum_{n=0}^{\infty}|c_{m,n}(\alpha)|^2={\textstyle\frac{1}{4}}
\left\{\sum_{n=0}^{\infty}|c^{(+)}_{m,n}(\alpha)|^2+
\sum_{n=0}^{\infty}|c^{(-)}_{m,n}(\alpha)|^2\right .
}
\nonumber \\ && \hspace{15ex} + \,
\left .
2 {\rm Re}\!\left[\sum_{n=0}^{\infty}
c^{(+)}_{m,n}(\alpha)c^{(-)}_{m,n}(\alpha)
\right]\right\} .
\label{B2}
\end{eqnarray}
Thus,
\begin{eqnarray}
{\cal N}_{m}^{(\pm)} =
\sum_{n=0}^{\infty}|c^{(\pm)}_{m,n}(\alpha)|^2
= 2{\cal N}_{m} - {\cal I}_{m} ,
\label{B3}
\end{eqnarray}
where ${\cal N}_{m}$ is given in Eq.~(\ref{1.3b}), and
\begin{eqnarray}
{\cal I}_m=\sum_{n=0}^{\infty}(-1)^{n+m}|c^{(\pm)}_{m,n}(\alpha)|^2 .
\label{B4}
\end{eqnarray}
Combining Eqs.~(\ref{B4}) and (\ref{Co3}) yields
\begin{eqnarray}
{\cal I}_m=\sum_{n=0}^{\infty}\frac{\Gamma^{2}(m+n+1)}
{\Gamma^{2}[(m+n+2)/2] n!}
\left(-\textstyle\frac{1}{2}|\alpha|\right)^{n+m}.
\label{B5}
\end{eqnarray}
Using the relation
\begin{equation}
\frac{\Gamma(2n)}{\Gamma(n+1/2)}
= \frac{4^n}{2\sqrt{\pi}}\,\Gamma(n),
\label{B6}
\end{equation}
from Eq.~(\ref{B5}) we obtain
\begin{eqnarray}
{\cal I}_m=\frac{1}{\pi}\sum_{n=0}^{\infty}
\Gamma^{2}\!\left[\textstyle\frac{1}{2}(m+n+1)\right]
\frac{(-2|\alpha|)^{n+m}}{ n!} .
\label{B7}
\end{eqnarray}
Substituting in Eq.~(\ref{B7}) for the $\Gamma$ function the
integral representation, we may write
\begin{eqnarray}
\lefteqn{
\hspace*{-5ex}
{\cal I}_m=\frac{(-2|\alpha|)^m}{\pi}
\int\limits_{0}^{\infty} dt_{1}
\int\limits_{0}^{\infty} dt_{2}
\bigg[
e^{-t_{1}}e^{-t_{2}}(t_{1}t_{2})^{(m-1)/2}
}
\nonumber \\ && \hspace{10ex} \times \,
\sum_{n=0}^{\infty}
\frac{(-2|\alpha|\sqrt{t_{1}t_{2}}\,)^{n}}{n!}
\bigg],
\label{B8}
\end{eqnarray}
which reduces to
\begin{eqnarray}
\lefteqn{
{\cal I}_m=\frac{(-2|\alpha|)^m}{\pi}
\int\limits_{0}^{\infty} dt_{1}
\bigg[
e^{-t_{1}}(t_{1})^{(m-1)/2}
}
\nonumber \\ && \hspace{10ex} \times \,
\int\limits_{0}^{\infty} dt_{2} \,
e^{-t_{2}}(t_{2})^{(m-1)/2} e^{-2|\alpha|\sqrt{t_1t_2}}
\bigg].
\label{B9}
\end{eqnarray}
We first calculate the $t_{2}$ integral \cite{Erdelyi1} to obtain
\begin{eqnarray}
\lefteqn{
{\cal I}_m=\frac{m!(-\sqrt{2}|\alpha|)^m}{\pi}
}
\nonumber \\ && \hspace{2ex} \times \,
\int\limits_{0}^{\infty} dt_{1} \,
2^{-(1-\alpha^2 t_1 /2)t_1}\,
(t_1)^{(m-1)/2}\,
{\rm D}_{-m-1}\!\left(\sqrt{2\alpha^2t_{1}}\right) .
\nonumber \\ &&
\label{B10}
\end{eqnarray}
Calculating the resulting $t_{1}$ integral \cite{Erdelyi1},
we eventually obtain
\begin{eqnarray}
\lefteqn{
{\cal I}_m
=\frac{m!^2(-|\alpha|)^m}{\sqrt{\pi}2^m\Gamma(m+3/2)}
}
\nonumber \\ && \hspace{5ex} \times \,
{\rm F}\!\left[\textstyle\frac{1}{2}(m\!+\!1),
\frac{1}{2}(m\!+\!1),m+\frac{3}{2},1-\alpha^2\right],
\label{B11}
\end{eqnarray}
where ${\rm F}(\alpha,\beta,\gamma,z)$ is the hypergeometric function.

The calculation of $\langle\Psi_{m}^{(\pm)}|\beta\rangle$
can be performed in a similar way. Using Eqs.~(\ref{1.4}),
(\ref{1.5}), (\ref{Q2}), and (\ref{B6}), we have
\begin{eqnarray}
\lefteqn{
\langle\Psi_m^{(\pm)}|\beta\rangle
=(\pm 1)^m\sqrt{\frac{(2\alpha)^m}
{\pi{\cal N}_m^{(\pm)}}} \, e^{-|\beta|^2/2}
}
\nonumber \\ && \hspace{10ex} \times \,
\sum_{n=0}^\infty
\frac{(\pm\sqrt{2\alpha}\beta)^n}{n!}
\Gamma\!\left[{\textstyle\frac{1}{2}}(n\!+\!m\!+\!1)\right] .
\label{B12}
\end{eqnarray}
Substituting in Eq.~(\ref{B12}) for the $\Gamma$ function again
the integral representation, we arrive at an integral of the type of
the $t_{2}$ integral in Eq.~(\ref{B9}), which may be
calculated using standard formulas \cite{Erdelyi1}.
We then obtain
\begin{eqnarray}
\lefteqn{
\langle\Psi_m^{(\pm)}|\beta\rangle
= (\pm 1)^{m}m!\sqrt{\frac{2\alpha^{m}}{\pi{\cal N}_m^{(\pm)}}}
}
\nonumber \\ && \hspace{10ex} \times \,
e^{-|\beta|^{2}/2+\beta^{2}\alpha/4} \,
{\rm D}_{-m-1}\!\left(\pm\sqrt{\alpha}\beta\right),
\label{B13}
\end{eqnarray}
where ${\rm D}_{m}(z)$ is the parabolic cylinder function.
\end{appendix}
%%%%%%%%%%%%%%%%%%%%%%%%%%%%%%%%%%%%%%%%%%%%%%%%%%%%%%%%%%%%

%%%%%%%%%%%%%%%%%%%%%%%%%%%%%%%%%%%%%%%%%%%%%%%%%%%%%%%%%%%%%%%%%%

\newpage

\newpage

%%%%%%%%%%%%%%%%%%%%%%%%%%%%%%%%%%%%%%%%%%%%%%%%%%%%%%%%%%%%%%%%%

\newpage

%%%%%%%%%%%%%%%%%%%%%%%%%%%%%%%%%%%%%%%%%%%%%%%%%%%%%%%%%%%%
\newpage

%%%%%%%%%%%%%%%%%%%%%%%%%%%%%%%%%%%%%%%%%%%%%%%%%%%%%%%%%%%%%%%

\newpage

%%%%%%%%%%%%%%%%%%%%%%%%%%%%%%%%%%%%%%%%%%%%%%%%%%%%%%%%%%%%%%%%

\newpage

\begin{thebibliography}{99}
%\bibitem[*]{Olomouc}
%Permanent address:
%Palack\'{y} University, Faculty of Natural Sciences, Svobody 26,
%77146 Olomouc, Czech Republic

\bibitem{Schroedinger1}
E. Schr\"{o}dinger,
Naturwissenschaften {\bf 23}, 807 (1935);
{\bf 23}, 823 (1935); {\bf 23}, 844 (1935).
 
\bibitem{Yurke1}
B. Yurke and D. Stoler,
Phys. Rev. Lett. {\bf 57}, 13 (1986).

\bibitem{Janszky1}
J. Janszky, A.V. Vinogradov, T. Kobayashi, and Z. Kis,
Phys. Rev. A {\bf 50}, 1777 (1994).

\bibitem{Walmsley1}
I.A. Walmsley and M.G. Raymer,
Phys. Rev. A {\bf 52}, 681 (1995).

\bibitem{Vogel1}
R.L. de Matos Filho and W. Vogel
Phys. Rev. Lett. {\bf 76}, 608 (1996).

\bibitem{Monroe1}
C. Monroe, D.M. Meekhof, B.E. King, and D.J. Wineland,
Science {\bf 272}, 1131 (1996).

\bibitem{Nieto1}
M.M. Nieto
Phys. Lett. A {\bf 219}, 180 (1996).

\bibitem{Buzek1}
V. Bu\v{z}ek and P. L. Knight, Progress in Optics XXXIV, Ed. E. Wolf
(North-Holland, Amsterdam 1995).

\bibitem{LaPorta1}A.
La Porta, R. E. Slusher, and B. Yurke,
Phys. Rev. Lett. {\bf 62}, 28 (1989).

\bibitem{Song1}
S. Song, C. M. Caves, and B. Yurke,
Phys. Rev. A {\bf 41}, 5261 (1990).

\bibitem{Yurke2}B. Yurke, W. Schleich and D. F. Walls,
Phys. Rev. A {\bf 42}, 1703 (1990).

\bibitem{Tombesi1}
P. Tombesi and D. Vitali,
Phys. Rev. Lett. {\bf 77}, 411 (1996).

\bibitem{Tombesi2}
P. Goetsch, P. Tombesi, and D. Vitali,
Phys. Rev. A {\bf 54}, 4519 (1996).

\bibitem{Ban1}
M. Ban, Phys. Rev. A {\bf 49}, 5078 (1994).

\bibitem{Ban2}
M. Ban, J. Mod. Opt. {\bf 43}, 1281 (1996).

\bibitem{Yurke3}
B. Yurke, S.L. McCall, and J.R. Klauder,
Phys. Rev. A {\bf 33}, 4033 (1986).

\bibitem{Campos1}
R.A. Campos, B.E.A. Saleh, and M.C. Teich,
Phys. Rev. A {\bf 40}, 1371 (1989).

\bibitem{Mandel1}
L. Mandel, Opt. Lett, {\bf 4}, 205 (1979).

\bibitem{Vogel2}
W. Vogel and D.-G. Welsch,
{\em Lectures on Quantum Optics}
(Akademie Verlag, Berlin, 1994).

\bibitem{Gradshteyn1}
I.S. Gradshteyn and I. M. Ryzhik, {\it Tables of
Integrals, Series and Products} (Academic, New York, 1980).

\bibitem{Walker1}
N.G. Walker,
J. Mod. Opt. {\bf 34}, 15 (1987).

\bibitem{Zucchetti1}
A. Zucchetti, W. Vogel, and D.-G. Welsch,
Phys. Rev. A {\bf 54}, 856 (1996).

\bibitem{Walker2}
N.G. Walker and J.E. Caroll,
Electron. Lett. {\bf 20}, 981 (1984).

\bibitem{Freyberger1}
M. Freyberger, K. Vogel, and W.P. Schleich,
Phys. Lett. A {\bf 176}, 41 (1993).

\bibitem{Paul1}
H. Paul, P. T\"orm\"a, T. Kiss, and I. Jex,
Phys. Rev. Lett. {\bf 77}, 2446 (1996).

\bibitem{Erdelyi1}
A. Erdelyi, {\it Higher Transcendental Functions}, 
Batman Manuscript Project, Vol 3
(McGraw-Hill, New York, 1953).



\end{thebibliography}
\end{document}